\documentclass[usenatbib,useAMS]{mn2e}
\usepackage{epsfig,psfig,graphicx,float,color}
\usepackage{subfigure}
\usepackage{mybibdefs}
\usepackage{multirow}
\usepackage{widetext}
\usepackage{url}
 \usepackage{color}
\usepackage{amsmath}

\newcommand{\bra}{\langle}
\newcommand{\ket}{\rangle}

\newcommand\rr{\color{black}}

\newcommand\bh{\color{black}}

\usepackage{graphicx}
\usepackage{amssymb}
\usepackage{epstopdf}

\title[Stacking for machine learning redshifts applied to SDSS galaxies]{Stacking for machine learning redshifts applied to SDSS galaxies}  
\author[Zitlau et al.]{Roman Zitlau$^{1}$, Ben  Hoyle$^{1}$, Kerstin Paech$^{1}$, Jochen Weller$^{1,2,3}$
 \newauthor 
Markus Michael Rau$^{1,3}$,  Stella Seitz$^{1,3}$\\\\
$^1$Universitaets-Sternwarte, Fakultaet fuer Physik, Ludwig-Maximilians Universitaet Muenchen, Scheinerstr. 1, D-81679 Muenchen, Germany\\
$^2$Excellence Cluster Universe, Boltzmannstr. 2, D-85748 Garching, Germany\\
$^3$Max Planck Institute fuer Extraterrestrial Physics, Giessenbachstr. 1, D-85748 Garching, Germany\\
\\
{\tt E-mail: rzitlau@usm.lmu.de, roman.zitlau@yahoo.de, benhoyle1212@gmail.com}
 }
  
\begin{document}
\date{Accepted ----. Received ----; in original form ----.}
\pagerange{\pageref{firstpage}--\pageref{lastpage}} \pubyear{2010}
\maketitle
\label{firstpage}
\begin{abstract}
We present an analysis of a general machine learning technique called `stacking' for the estimation of photometric redshifts. Stacking techniques can feed the photometric redshift estimate, as output by a base algorithm, back into the same algorithm as an additional input feature in a subsequent learning round. We shown how all tested base algorithms benefit from at least one additional stacking round (or layer). To demonstrate the benefit of stacking, we apply the method to both unsupervised machine learning techniques based on self-organising maps (SOMs), and  supervised machine learning methods based on decision trees. We explore a range of stacking architectures, such as the number of layers and the number of base learners per layer. Finally we explore the effectiveness of stacking  even when using a successful algorithm  such as AdaBoost. We observe a significant  improvement of between 1.9\% and 21\% on all computed metrics when stacking is applied to weak learners (such as SOMs and decision trees).  When applied to strong learning algorithms (such as AdaBoost) the ratio of improvement shrinks, {\bh but still remains positive and is between 0.4\% and 2.5\% for the explored metrics and comes at almost no additional computational cost.}
\end{abstract}
\begin{keywords}
galaxies: distances and redshifts,  catalogues, surveys.
\end{keywords}

\section{introduction}
The advent of massive photometric surveys of the night sky has created a data-rich playground in which our knowledge and assumptions about theoretical astrophysics and cosmology can be critically tested. Photometric surveys are most useful for cosmological analysis once stars, galaxies and artefacts have been correctly identified, and their properties and positions measured. 

The most accurate measurement of redshift for galaxies identified in such surveys is through spectroscopic follow up on large telescopes. However this is a time consuming endeavour, and can only be performed for a subset of all of the photometrically identified galaxies. The challenge is then to attempt to estimate redshifts for the full photometric sample using either our knowledge of the evolution of stellar population models, which is  encoded in template fitting routines, or to use the subset of galaxies with spectra to learn a mapping between the measured photometric properties (or `features') of the galaxy and the redshift. These methods can then be applied to all representative photometric galaxies to estimate a redshift. If the photometric sample is not representative, techniques from machine learning, such as the covariate shift \citep[see e.g.][]{2014MNRAS.445.1482S,RauEtAllinPrep} or data augmentation \citep[see][]{2004A&A...423..761V,2015arXiv150106759H} can still help to obtain accurate redshift estimations. One can further aid this mapping process by optimally selecting which targets to follow up spectroscopicly \citep[][]{2015ApJ...813...53M,2015arXiv150806280H}.

Machine learning techniques exploit the wealth of existing photometric and spectroscopic data to provide the most accurate photometric redshift measurements available. As such, many different machine learning algorithms have been applied to the problem of estimating photometric redshifts and errors, and to estimate the full redshift distribution function \citep[e.g., ][]{2003LNCS.2859..226T,2004PASP..116..345C,2007AN....328..852C,2008ASPC..394..521C,2010ApJ...715..823G,2013arXiv1312.1287B,tpz,2015MNRAS.449.2040H,RauEtAllinPrep,2015arXiv150407255H}.

In this paper we apply the machine learning architecture called `stacking' to the problem of photometric redshift estimation. A stacked architecture passes the output of one machine learning algorithm as an additional input into a subsequent training round. Each training round, or `layer', may contain one or more algorithm and the algorithms may differ between training rounds. 

The concept of stacking is related to the combination of different predictions from many different systems, for example by using Baysean model combination for photometric redshift anaysis \citep{2014MNRAS.442.3380C}, or for star \& galaxy separation  \citep{2015MNRAS.453..507K}. Our work differs from Baysean model combination approaches by passing both the photometric features, {\it and} the algorithm predictions into similar algorithms in a subsequent training layer. We then explore the effect of having many consecutive training layers to find the best balance between layers, and the number of algorithms which are combined per layer.

There are analogies between stacking architectures and deep machine learning \citep[][]{lecun95convolutional,NIPS2012_4824,2015arXiv150307077D,2015arXiv150407255H}, if one considers that each layer of the deep neural network could learn a model and pass the model prediction to a deeper layer along with the original, albeit rescaled, input features.

The format of the paper is as follows. We describe the data sample in \S\ref{data}, and continue by describing the machine learning algorithms and  stacking architectures in \S\ref{method}. We describe the analysis and present results in \S\ref{results}, discuss in \S\ref{discussion}, and conclude in \S\ref{conclusions}.

\section{Data}
\label{data}
The data in this study are drawn from Sloan Digital Sky Survey Data Release 10 \citep[hereafter SDSS,][]{2014ApJS..211...17A}. The SDSS I-III uses a 4 meter telescope at Apache Point Observatory in New Mexico and has CCD wide field photometry in 5 bands \citep[$u,g,r,i,z$][]{Gunn:2006tw,Smith:2002pca}, and an expansive spectroscopic follow up program \citep[][]{2011AJ....142...72E} covering $\pi$ radians of the northern sky. { The SDSS collaboration has obtained approximately 2 million galaxy} spectra using dual fiber-fed spectrographs. An automated photometric pipeline performed object classification to a magnitude of $r\approx$22 and { measured} photometric properties of more than 100 million galaxies. The complete data sample is publicly available through the CasJobs server \citep[][]{10.1109/MCSE.2008.6}\footnote{skyserver.sdss3.org/CasJobs}.
We select 1,958,727 galaxies from the CasJobs server with both spectroscopic redshifts and photometric properties. In detail we run the MySQL query shown in \S\ref{mysql}, within the DR10 schema context.

\subsection{Input photometric features}
\label{featscal}
For this work, we follow \cite{2014A&A...568A.126B} and select the five band $u,g,r,i,z$ psf magnitudes to create input features for the machine learning algorithms. We apply extinction corrections to the psf magnitudes, and further only select galaxies that have a photometric galaxy classification $type=3$. We further remove spurious objects by selecting objects which have psf magnitudes between 10 and 30. The maximum allowed measurement error is set to  0.6 (0.5) for the psf magnitude $u$ ($g$) and 0.1 for $r, i$ and  $z$.  We note that this choice of photometric features is rather arbitrary, and recent work has shown how a machine learning inspired selection of photometric features, can lead to improvements for redshift estimation \citep{2014arXiv1410.4696H}.

In this work we use the spectroscopic redshift as the output feature for the machine learning algorithms and therefore apply cuts to this feature very carefully. We only remove data with spectroscopic redshifts smaller than zero, because they are not physical, and further remove objects with a spectroscopic redshift error greater than 0.0002. This reduces the sample size to 1,134,080 galaxies. 
We have checked that these cuts do not lead to significant differences in the input feature space of the data set. 


\subsection{Training, validation and test samples}
In this work we construct and train many hundreds of machine learning systems. To keep the training times manageable, we follow \cite{2014MNRAS.442.3380C}, and randomly sub-divide the input catalogue into training, validation and test samples of size 50k, 200k and 884k respectively. This corresponds to approximately 4 \%, 18\% and 78\% of the whole data set. The training sample is always used to train the machine learning algorithm for a given hyper-parameter set and stacking architecture. We use the validation sample to select good hyper-parameters values for the machine learning algorithms. After the selection process is finished, we pass the independent test sample through the  machine learning systems to measure the performance. This ensures that the results as measured on the test sample describe the ability of the machine learning system to generalise to new, albeit representative, datasets.

\section{Machine learning methods}
\label{method}
In this section we present a short overview of different supervised and unsupervised learning approaches. We introduce and describe the machine learning algorithms such as self-organising maps (hereafter SOM), decision trees, and AdaBoost, and describe their hyper-parameters. We then introduce ensemble learning methods like bagging, boosting, and finally describing stacking and the stacking architectures explored in this work. Ensemble methods combine the predictions obtained from many base learners, such as one SOM, or one decision tree, and can reduce both bias and variance of predictions obtained when using just one base learner.


\subsection{Self-Organising Maps}
{\bh
A SOM \citep{Kohonen:2001:SM:558021} is an unsupervised artificial neural network algorithm, which is capable of approximating non-linear functions of arbitrary complexity hidden within the data, by clustering the data onto neurons. Each neuron can be represented graphically as a map, and then each neuron represents a cell in the map. The amount of input data and the number of neurons of the SOM are correlated with the number of objects which are assigned to each cell. Depending on the chosen topology of the SOM, each cell can have a different number of direct or nearby neighbours, which influence the predicted values. 

The SOM is initially constructed with randomly selected weights connecting the input data to the neurons. The photometric properties of the training is passed through the SOM and the value of the neuron which best represents the galaxy is defined as the winning neuron. The weights of the winning neuron and its neighbouring neurons are adjusted to make the neuron even more similar to this training example. The magnitude of the weight updates for the surrounding neurons decreases as the distance on the map to the winning neuron increases. The training dataset is presented to the SOM many times in a random order, during an iterative training process. This results in a grouping of similar data around nearby neurons. The goal of this learning is the creation of a two-dimensional representation of the multidimensional data, which can reveal similarities and correlations in the data. During the training phase the spectroscopic redshift information is not used in the construct of the SOM. Finally the training data is passed through the SOM and the final cell that each galaxy belongs to is determined. The average spectroscopic redshift of all galaxies in each cell is calculated and assigned to the cell. New test data, which was not used during training, is passed through the SOM and lands in a cell. The cell redshift becomes the predicted photometric redshift assigned to the galaxy.
}

{\bh We choose a two-dimensional spherical topology with an online learning algorithm of 100 iterations. The SOM redshift predictions for new data are assigned using the mean of the spectroscopic redshifts of all training galaxies belonging to the winning neuron and its neighbours. There are several other parameters controlling the SOM architecture, especially the learning and prediction processes. Further hyper-parameters, such as the learning rates, are fixed following  \citep{2014MNRAS.438.3409C}, and we refer the reader to this work for a detailed description of all of the hyper-parameters of a SOM.
}

We build upon the work of \cite{2014MNRAS.438.3409C}, in which many SOMs are combined to form a  `Random Atlas' which can be applied to photometric redshift estimation. In this work we also employ the same publically available Fortran implementation of SOMz\footnote{lcdm.astro.illinois.edu/code/mlz.html}.  We further choose to fix the geometry of the SOM to be spherical, and adjust the hyper-parameter   $n$ of the SOM which results in the number of neurons per SOM being $12\times n\times n$.

\subsection{Decision Trees}
A decision tree is a supervised machine learning algorithm, which recursively partitions the high dimensional input feature space, into an increasing number of hyper-cubes, or `leaf nodes'. New and smaller subsets (or `branches') of the data are created until only a specified minimum number of objects are grouped onto a final leaf node. The algorithm selects a feature dimension and decides at which point to split the data into two cubes by trying to minimise the mean squared error in the output feature of the data in each sub cube.  After training is finished the test data is queried down the tree and lands on a leaf node. The training galaxies in that  leaf node are used to produce redshift predictions for the test data. For a more detailed description of the decision tree algorithm we refer the reader to \cite{ig}, and for possible hyper-parameter choices we refer the reader to the  scikit-learn \citep[][]{scikit-learn} documentation.

We note that constructing a decision tree with millions of galaxies, and many tens of feature dimensions is achieved in just tens of seconds, whereas a comparable sized SOM requires tens of hours. One single decision tree is often referred to as a `weak learner' because it is only capable of learning  either a very simple model, with little predictive power, or a very complex model which will be prone to over fitting.

The hyper-parameter of the decision trees that is explored in this work is the minimum the number of training examples which sit upon each final leaf node, all other parameters are fixed to the default values set by the scikit-learn routines.

\subsection{Bagging ensemble learning frameworks}
\label{ens}

The concept of ensemble learning \citep{Dietterich:2000:EMM:648054.743935} is based on the assumption that a collection of weak learners, such as one decision tree, or one SOM, can be combined into one ensemble system, which is a strong learner. Strong learners have a lot of predictive power and suffer less from over-fitting.

One common method to produce an ensemble of weak learners is called a `Random Forest' \citep{RandoMforests} which commonly combines many decision trees, and an `Atlas of SOMs' \citep{2014MNRAS.438.3409C} when combining SOMs. Both of these algorithms construct ensembles of training data with which to train each weak learner independently. The ensembles of training data are selected by `bagging'. Bagging creates many new training datasets by randomly sampling from the original training dataset with replacement. The re-sampling process can also include a round of perturbing the data set by the measurement errors, which is a form of data augmentation \citep[see also][]{tpz,2015arXiv150106759H}. Bagging is not restricted to the training examples, but may also be used to select a subset of the input photometric features presented to each weak learner.

In this work we use bagging by selecting from both training data, and input features, and by also applying a Gaussian re-sampling of the feature values from the corresponding feature errors. We refer to this combined process as `bootstrap re-sampling'.

{\rr In this work we explore the number of learners as the hyper-parameter of both the random forest algorithm, and the random atlas algorithm which sit on each `stacked layer'. However the entire stacked system is not a normal random atlas or random forest, see \S\ref{as}, for more details. }

\subsection{Boosting ensemble learning frameworks}
\label{adab}
Adaptive Boosting, or AdaBoost \citep[][]{Freund1997119,Drucker:1997:IRU:645526.657132} is an  ensemble learning algorithm which is often uses decision trees as the base learning. Each learner is trained in sequence. The first learner constructs a dataset by bootstrap resampling from the features space, and possibly the data, and trains a model which makes a base prediction. Each following learner again randomly samples from the feature space but concentrates the training process on those training examples which were poorly estimated in the previous round. The  AdaBoost algorithm therefore learns from previous training rounds, in order to concentrate on the most difficult examples. This technique produces a stronger learner which often has more predictive power than other ensemble methods, such as random forests \citep[e.g.,][]{2015MNRAS.452.4183H}. 

There are other boosting algorithms, such as Gradient Boosting \citep{friedman2001} which only applies weights to each successive weak learner, and not the data set. We have found that Gradient Boosting performs worse than AdaBoost for the task of photometric redshift estimation.

In this work, the explored hyper-parameters of the AdaBoost algorithm are the number of decision tree learners, and number of training examples which sit on each decision tree leaf node.

\subsection{Stacking}
\label{as}

\begin{figure*}
\includegraphics[scale=0.85, clip=true, trim=15 275 40 35]{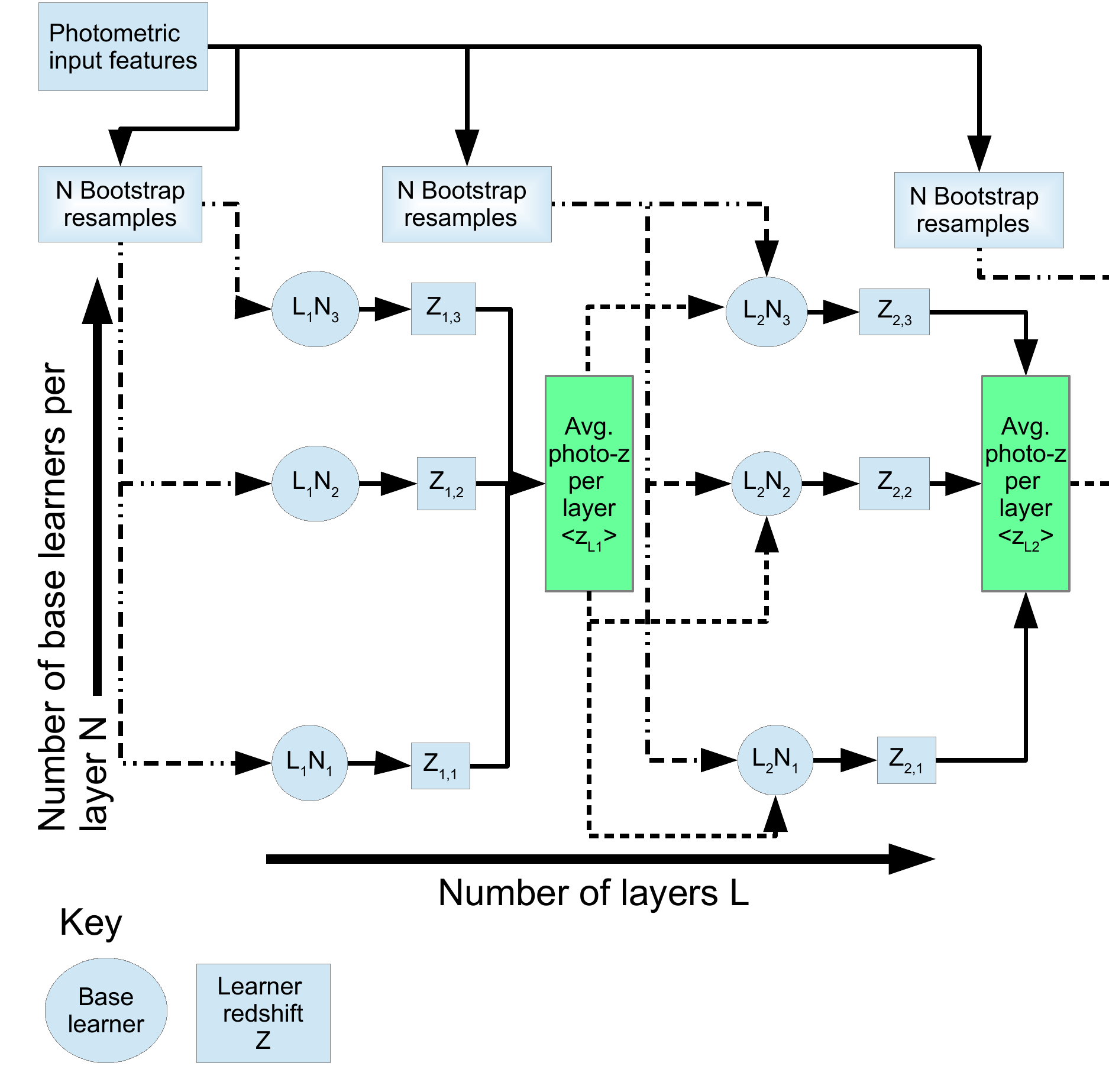}
\caption{ {\bh A schematic diagram of the stacking procedure. Each stacking layer has $N$ learners at each `$i^{th}$' layer $L_i$, and each learner is given a bootstrapped re-sampled training set. The $N$ learners are trained and then each produce a redshift estimate for each training galaxy, as denoted by $Z_{i,j}$. This is achieved by passing all training data through layer $L_i$ and obtaining redshift estimates from each of the $N$ learners individually. We compute the mean of all $N$ redshift values for each training example, as denoted by $\bra z_{L_i}\ket$. The photometric data is again bootstrap re-sampled $N$ times, and then the feature values and the mean redshift value $\bra z_{L_i}\ket$ for each galaxy is passed into each of the $N$ learners on the next layer $L_{i+1}$.} }
\label{flowchart}  
\end{figure*}

The purpose of this paper is to explore the concept of `stacking' \citep{Wolpert92stackedgeneralization}  and apply it to the problem of photometric redshift estimation. Stacking is similar to the above ensemble algorithms in that it incrementally constructs sets of learners. In this work we explore both the use of weak learners, and strong learners, as the base learner. Stacking is different from boosting and bagging because each layer $L$, is used to produce an actual redshift estimate for each galaxy, and this estimate is included as an {\it additional} input feature into the next layer of the system.

There are many potential ways to pass each new redshift estimate into subsequent layers, and we have explored the following {\rr five} methods. 1: We propagate only one redshift estimate chosen to be; 1.a: the mean (1.b: the median) of predictions from all $N$ learners on that layer, as an additional feature into the next layer, and discard all redshift estimates from earlier layers. This increases the input feature dimension from the base number of photometric features $N_{PF}$, to $N_{PF}+1$ at every additional layer.  2.a: We have also explored passing the mean (2.b: the median) photometric redshift estimates from the average values of all previous layers as a new input feature to the next layer, this again keeps the input feature dimension fixed to $N_{PF}+1$.  3: We also cumulatively propagate every mean redshift estimate for all preceding layers into the subsequent layer, thereby increasing the number of features by $N_{PF}+L$ for the $L^{th}$ additional layer. From this collection of possibilities we concentrate on the approach which performs the best on the  validation sample, which is approach 1.a. We note the approach 1.b improves the median metric value more than approach 1.a, which however improves the other two metrics (which are described further in \S\ref{metrics}).

{\bh 
We provide a schematic diagram of the stacking architecture in Fig. \ref{flowchart}, and describe the stacking procedure further with the aid of this diagram. Each stacking layer has $N$ learners at each `$i^{th}$' layer $L_i$, and each learner is given a bootstrapped re-sampled training set. The $N$ learners are trained and then each produce a redshift estimate for all training data, as denoted by $Z_{i,j}$ in Fig. \ref{flowchart}. This is achieved by passing all training data through layer $L_i$ and obtaining redshift estimates for each galaxy from each of the $N$ learners. We compute the mean of all $N$ redshift values for each training example, as denoted by $\bra z_{L_i}\ket$ in the schematic diagram. The photometric data is again bootstrap resampled $N$ times, and then the feature values and the mean redshift value $\bra z_{L_i}\ket$ calculated for each galaxy, is passed into each of the $N$ learners on the next layer $L_{i+1}$.
}

We also compare the effect of stacking, against simply using a similarly larger number of learners in the ensemble. This means that for every stacking system which has $L$ layers, and $N$ learners per layer, we also create an ensemble learner with $L\times N$ base learners which does not perform stacking. In the cases that a strong learner, such as AdaBoost with $M$ decision trees, is used as a base learner in the stacking system, we correspondingly increase the number of base learners in the comparison system by a factor $M$, thereby bringing the total number of base learners to $L \times N \times M$.

We explore different combinations of $L$, and $N$ (or $L$, $N$ and $M$) such that the total number $P = L\times N$ (or $P=L\times N\times M$) of base learners is fixed, for example we explore $P=L\times N =1\times 100$, $P=2\times 50$, $P=4\times 25$, ..., $P=1\times 100$. We also explore different values of $P$.

We note that $N=1$ correspondes to having only one learner per layer and $L=1$ represents having no stacking layer, and therefore corresponds to the comparison systems. We explore a grid of values, $P$, $L$, $N$ ($M$). For each of the values, we construct many realisations of the chosen system by randomly exploring the hyper-parameter space of the learners. We choose the realisation which performs the best on the validation set, as the system to use for those values of $P$, $L$, $N$ ($M$). One could consider these values as the hyper-parameters of the stacking architecture.

Finally we note that a linear combination of many decision trees is the definition of a random forest \citep[][]{RandoMforests}, in which the predictions of each of ${\rm N_t}$ trees $z_{\rm tree, i}$ are averaged to produce a redshift $ z_{ \rm RF}$ following

\begin{equation}
   z_{ \rm RF}  = \frac{1}{\rm N_t} \sum z_{{\rm tree, i}} \thinspace .
\label{eq:pred_reg_tree}
\end{equation}

In this work we do not use the term random forest because some stacking architectures have only one decision tree per layer, and many layers, and therefore these collections are not random forests. For the cases when there are many decision tree learners per layer, then each layer can be thought of as being a random forest, however the complete system is not a standard forest, and we therefore prefer to use the less misleading term `collection of decision trees'.

\section{Analysis and Results}
\label{results}

\begin{table}

\centering
\renewcommand{\footnoterule}{}
\begin{tabular}{ll}
Metric & Description\\
\hline
\hline
$|\Delta z^{50}|$ & the absolute value of the median of $\Delta z$.\\
$\sigma_{68}$ & The dispersion at which 68\% of $\Delta z$ is \\
\ & enclosed symmetrically around the median.\\
${\rm out }_{0.15}$ & Fraction of outliers {\rr with $|\Delta z| > 0.15$}\\
\end{tabular}
\caption{The definition of the metrics used to compare different machine learning performances, as measured on the redshift scaled residuals, defined as $\Delta z = (z_{\rm phot}-z_{\rm spec})/(1+z_{\rm spec})$.}
\label{tab:def_metrics}
\end{table}

\label{metrics}
The performance of the different stacking configurations are compared by predicting redshifts for the test data, which is not used during training. The redshift scaled residual vector is construct from the redshift predictions and the spectroscopic redshifts, and is defined as $\Delta z = (z_{\rm phot}-z_{\rm spec})/(1+z_{\rm spec})$. We choose to measure the following performance metrics on $\Delta z$; the absolute value of the median value of $|\Delta z^{50}|$, the 68\% spread of $\Delta z$ denoted by  $\sigma_{68}$, and the fraction of outliers ${\rm out}_{0.15}$, defined as the faction of predictions for which  $|\Delta z|>0.15$. We present and describe these metrics in Table \ref{tab:def_metrics} to aid the reader. {\bh We do not choose to compare other performance metrics, for example those which compare the full shapes of the predicted and true redshift distributions \citep[][]{2014MNRAS.445.1482S,RauEtAllinPrep,2015arXiv150705909B}, which can be useful metrics depending on the science case at hand.} In this paper we identify improvement if all of the three specified metrics are improved. This is similar to the `I-metric' weighting proposal presented in \citep{2014MNRAS.442.3380C,2014MNRAS.438.3409C}.

To generate error estimates on the measured metric quantities, we perform the following two sets of analysis. In all of the sections except in \S\ref{summarytest}, we train three machines for each set of machine hyper-parameters and stacking configurations, using three bagged re-samples of the training set. We then pass the  validation sample through each of these three machines and measure the metrics. The mean and standard deviation of these three values are presented in the figures, and we choose the best hyper-parameter configuration from these runs. In \S\ref{summarytest} we are more interested in how these results generalise to new data. We therefore choose the best hyper-parameter values using the validation sample, and run 10 new machine systems with these hyper-parameter values each having a bagged re-sampled training set. We pass the large sample of test data through each of the 10 systems and measure the metric values. In \S\ref{summarytest} we report and present the mean and standard deviation of these 10 values, as measured on the test sample.
 
\subsection{Different stacking architecture configurations}
\label{grid_search}

In this section we explore the effect of different stacking configurations with sets of fixed machine learning architecture hyper-parameters. To this end we choose to follow \cite{2014MNRAS.438.3409C} and set the SOM architecture  to have 768 neurons. We likewise create a comparable sized decision tree by fixing the `minimum number of objects per leaf' to be 65, which approximately corresponds to the same number of training samples on each leaf node, as are assigned to each cell of the SOM. We next choose the following fixed total number of learners in the stack $P$, to be $( 49, 100, 200, 400)$ and explore a range of the number of layers $L$, and learners per layer $N$, such that the following condition is always satisfied  $P=L\times  N$. If exact combinations of $L,N$ are not possible for each $P$, we choose close by values.

\begin{figure*}
\includegraphics[scale=0.37, clip=true, trim=35 34 85 36]{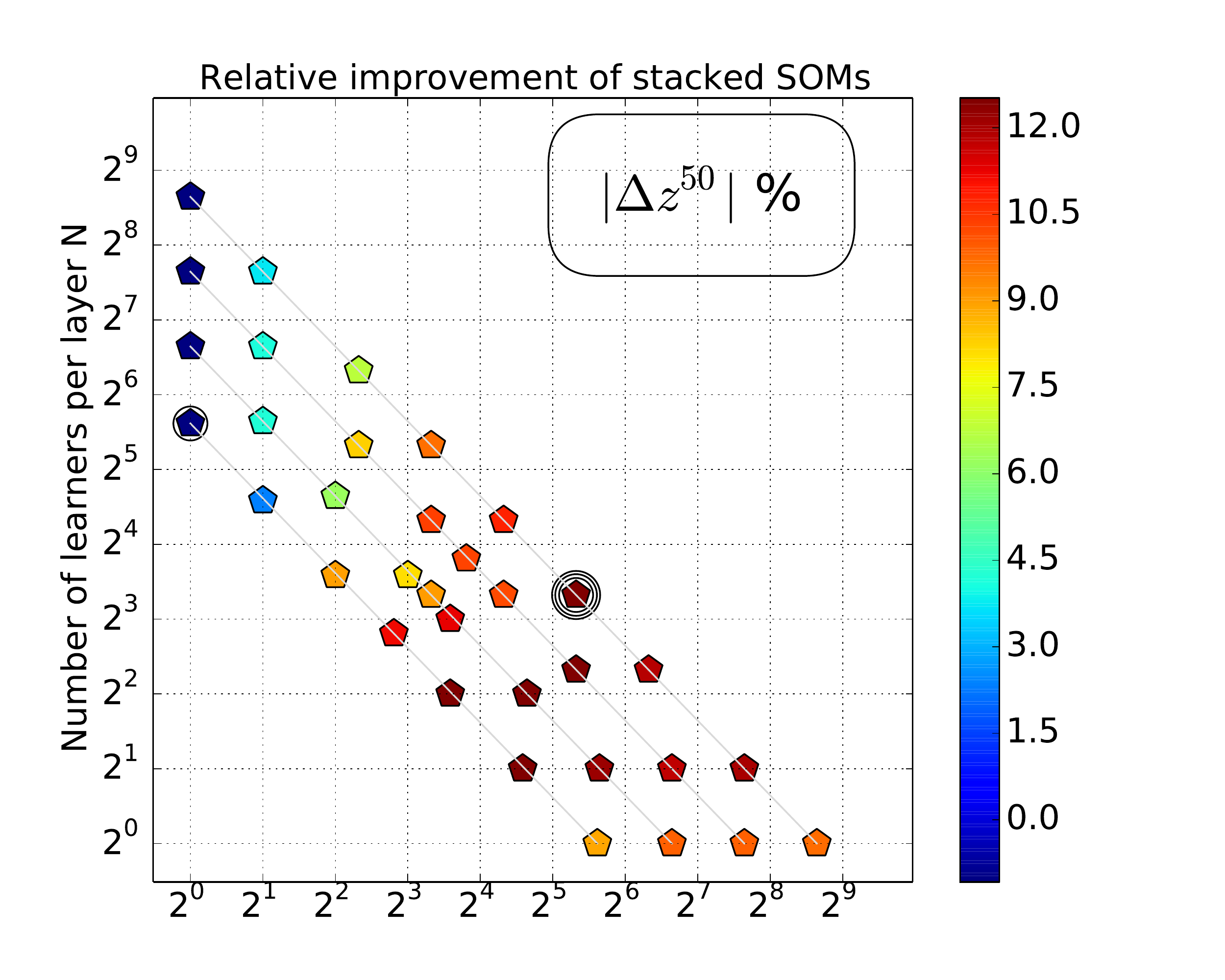}
\includegraphics[scale=0.37, clip=true, trim=35 34 70 36]{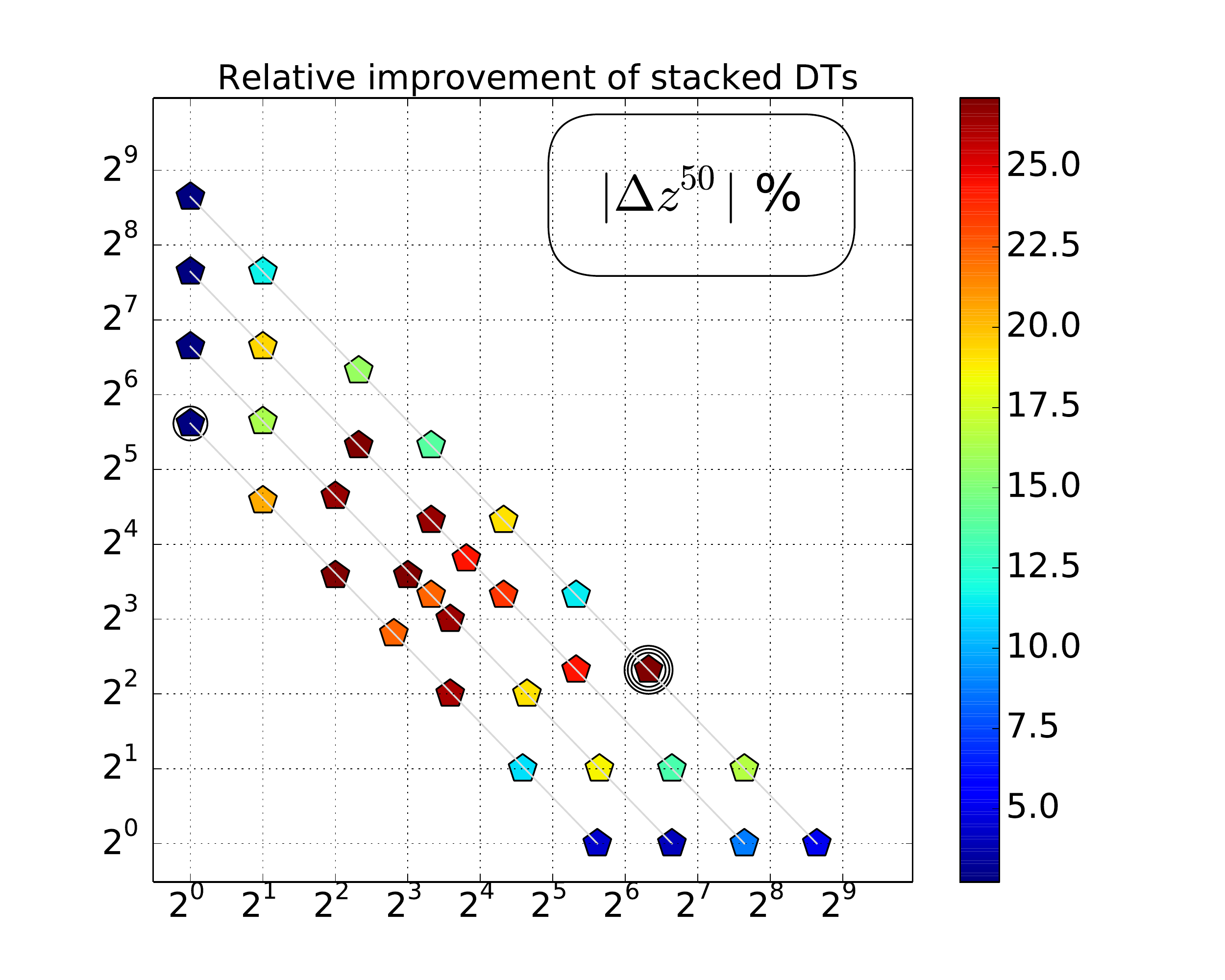}
\includegraphics[scale=0.37, clip=true, trim=35 34 85 44]{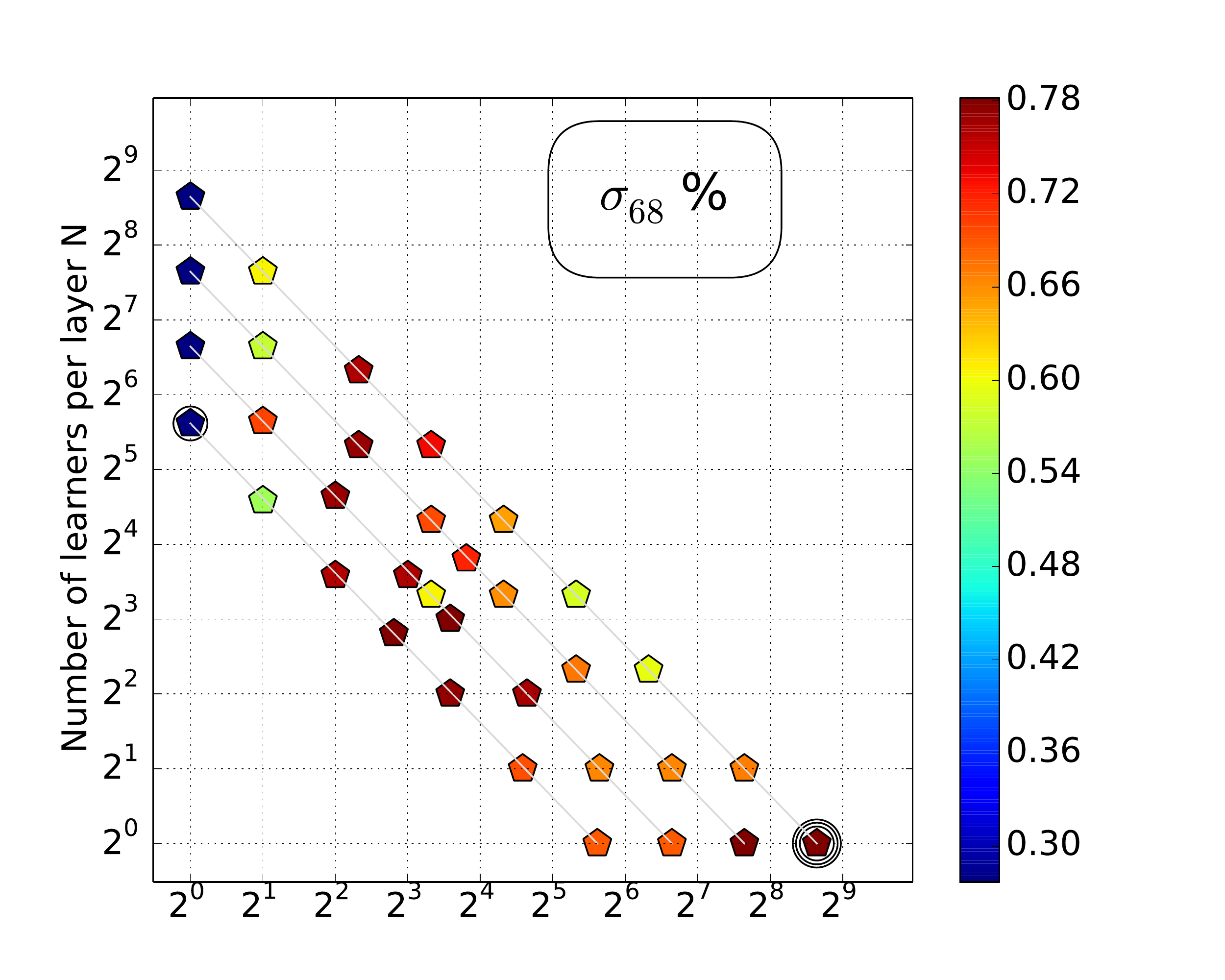}
\includegraphics[scale=0.37, clip=true, trim=35 34 70 44]{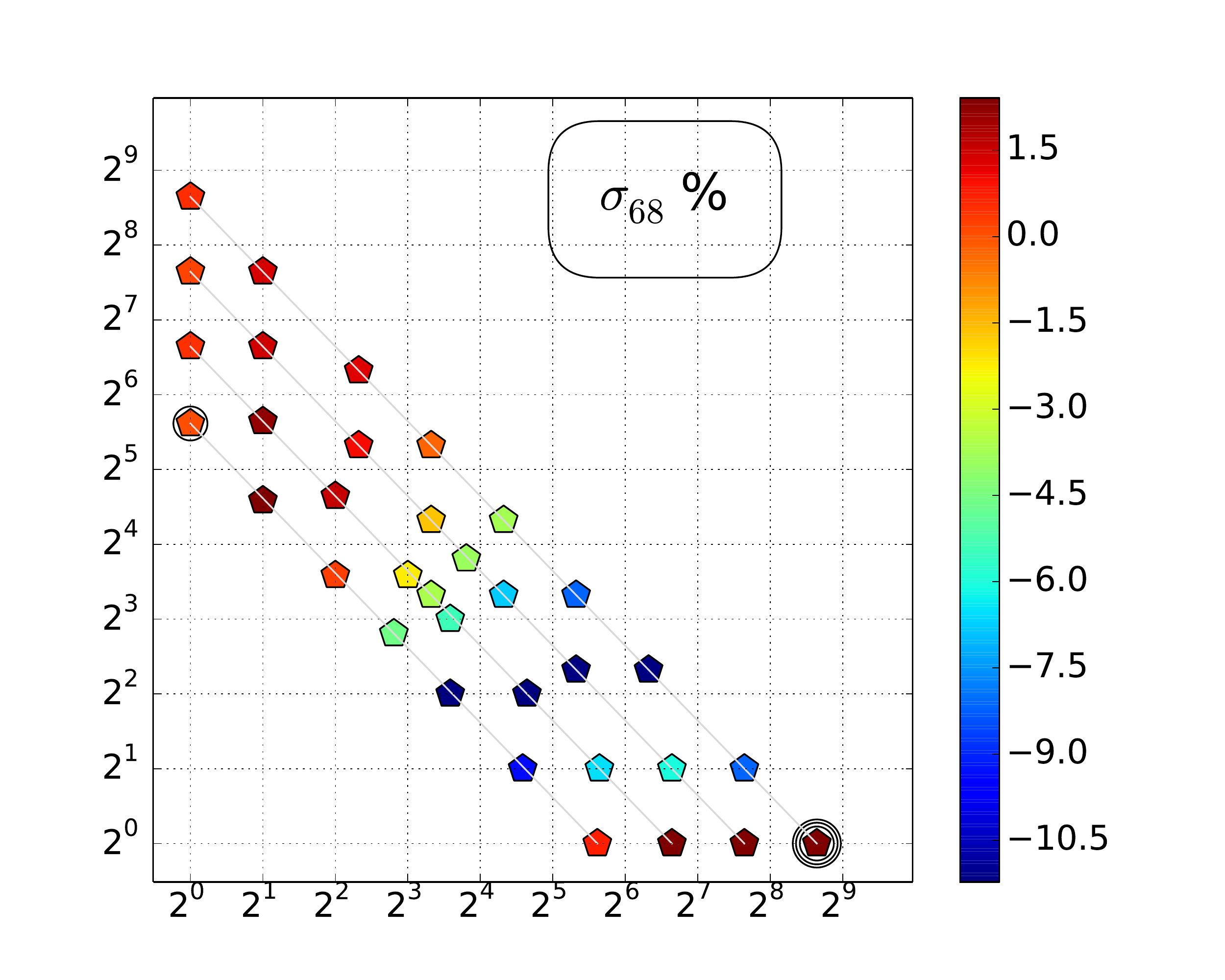}
\includegraphics[scale=0.37, clip=true, trim=35 2 85 44]{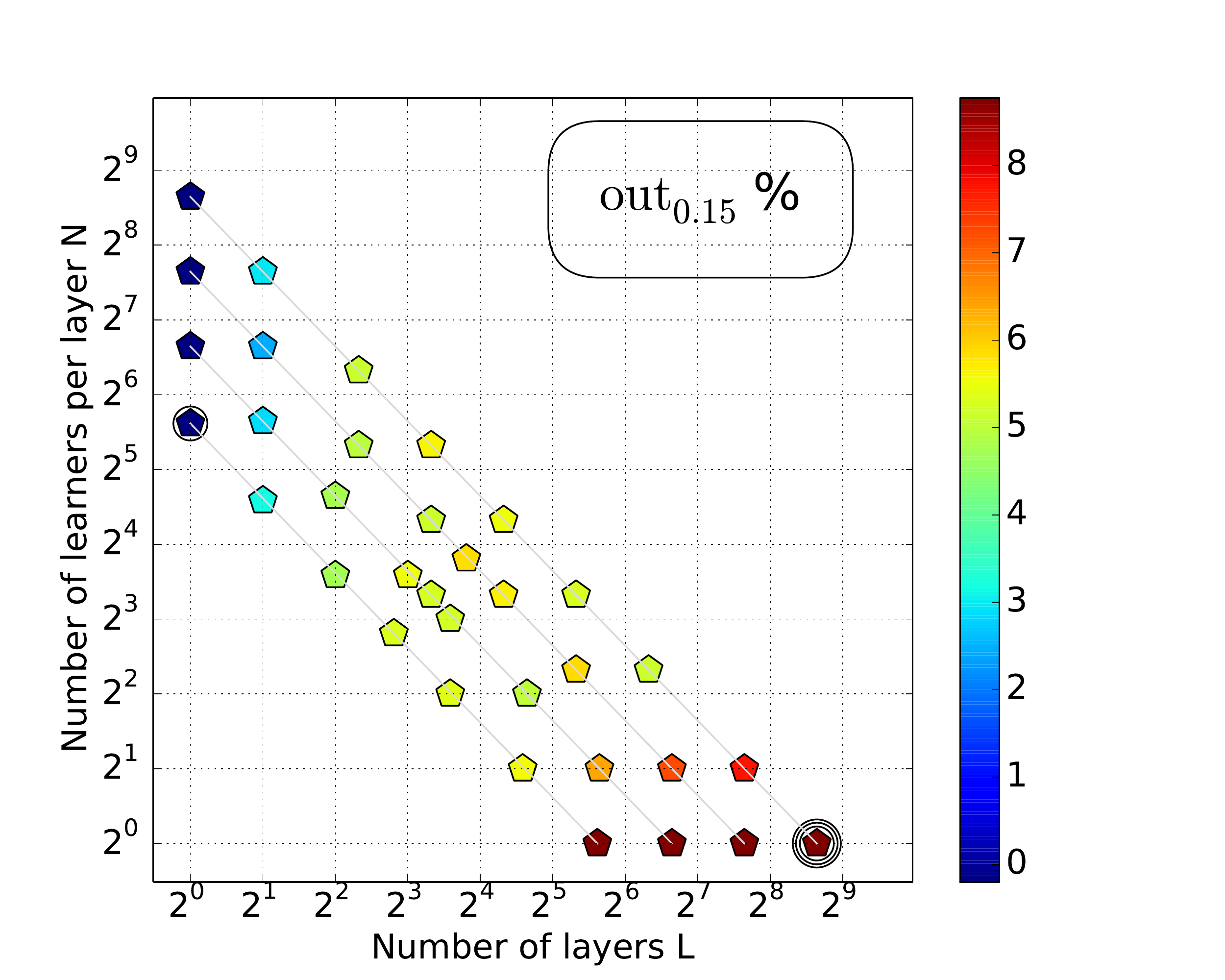}
\includegraphics[scale=0.37, clip=true, trim=35 2 70 44]{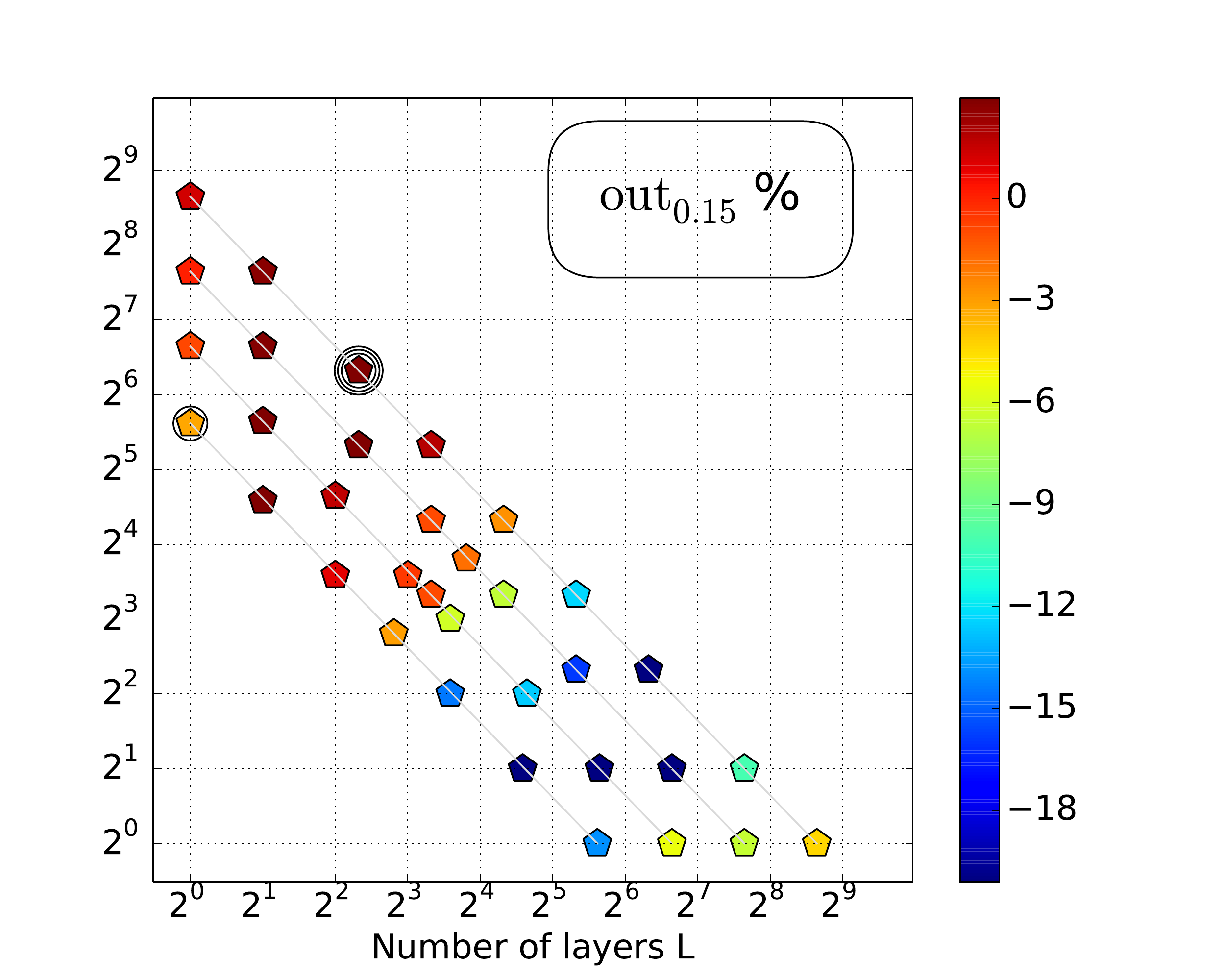}

\caption{ {\bh The {\rr{\it relative}} performance in each metric value as the stacking configuration changes.  The x-axis in each plot shows the number of stacking layers $L$, and the y-axis shows the number of learners per layer $N$. No stacking layers corresponds to $L=1$ ($\equiv 2^0$). The diagonal lines highlight constant values of $P=L\times N$. Redder (bluer) colors correspond to improvement (degradation) in the metric value as compared with the comparison system, which is highlighted by one circle.  The comparison system is that with no extra layers of stacking ($L=1$) and the smallest number of base learners $N$. The colors in each panel correspond to different ranges as shown in the legend. The left-hand (right-hand) panels show the effect of stacking using a SOM (decision trees) as the base learner. The top, middle and lower panels show the median value, $\sigma_{68}$, and the outlier fraction metrics respectively. The best overall system is further highlighted by three concentric circles.}  }
\label{gridsearchImg}  
\end{figure*}

{\bh
We present the results of this analysis in the panels of Fig. \ref{gridsearchImg}, which show the number of layers on the x-axis, and the number of learners per layer on the y-axis. Recall that $L=1$ ($\equiv 2^0$) indicates no extra stacking layers, and $L>1$ indicates extra layers of stacking. The diagonal lines on each panel show lines of fixed $P=L\times N=$constant. The left-hand panels show the effect of stacking using a SOM as the base learner, and the right-hand panels show the use of a decision tree as the base learner. The top, middle and lower panels show the median value, $\sigma_{68}$, and the outlier fraction metrics respectively, of the {\rr {\it relative} metric values, with respect to the comparison system}. Each data point corresponds to a particular choice of number of learners per layer, and the number of layers. Redder (bluer) colors correspond to improvement (degradation) in the metric value as compared with the comparison system, which is highlighted by one circle.  The comparison system is that with no extra layers of stacking ($L=1$) and the smallest number of base learners $N$. The colors in each panel correspond to different ranges as shown in the legend. The best overall architecture is highlighted by three concentric circles.
}

If we concentrate on each of the diagonal lines in each of the panels, we find that there is always a stacking configuration which improves each of the metric values when using both the SOM and decision trees as the base learner, as compared to the comparison system which does not have any stacking layers.  This can be seen by the change in color of the data points along each of the diagonal lines. In the left-hand panels of Fig. \ref{gridsearchImg} we see that the relative improvement in metric values continues to increase in all metrics as more and more stacking layers are added. For the outlier fraction metric this improvement is very clear as shown in the bottom panel. The panels corresponding to $\sigma_{68}$ and the median value also improve with the addition of extra stacking layers, however there is often a saturation level, after which increasing the number of stacking layers, and therefore decreasing the number of learners per layer, results in a degradation of results. This degradation is probably due to the lower number of learners per layer over-fitting the data and thereby reducing the predictive power of the machine learning system. 

By concentrating on the right-hand panels of Fig. \ref{gridsearchImg} we find that the metric values of ensembles of decision trees are also improved when at least one stacking layer is used, however the exact number of stacking layers which provide the best metric values is not the same for each metric, the outlier fraction performs better with more layers and less learners per layer, and the median value performs better with approximately equal numbers of layers and learners per layer.

\begin{figure}
\includegraphics[scale=0.29, clip=true, trim=8 14 90 4]{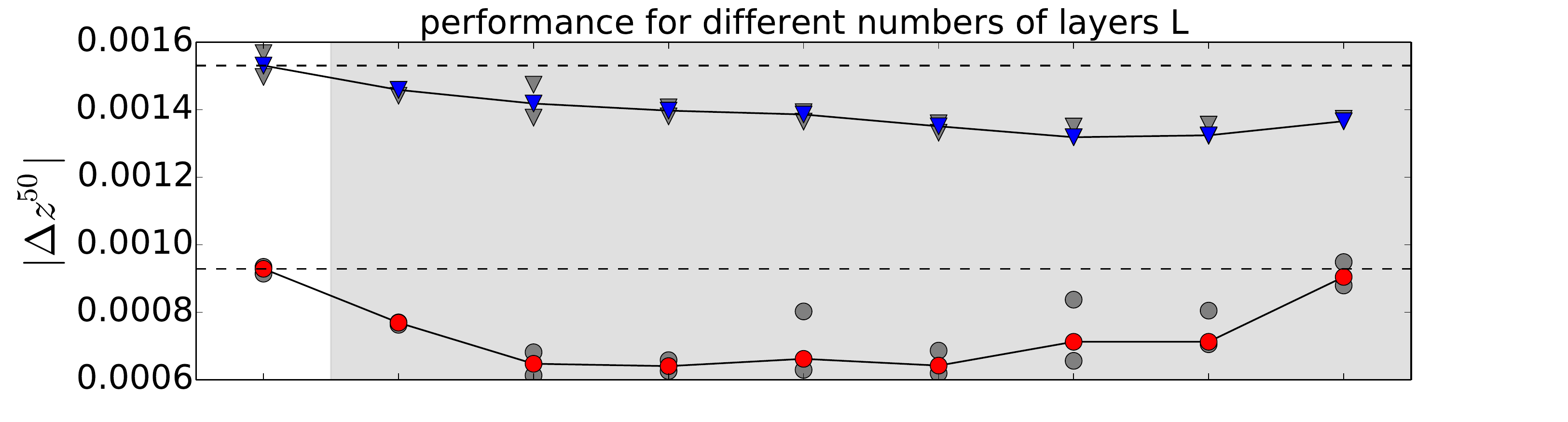}
\includegraphics[scale=0.29, clip=true, trim=8 14 90 16]{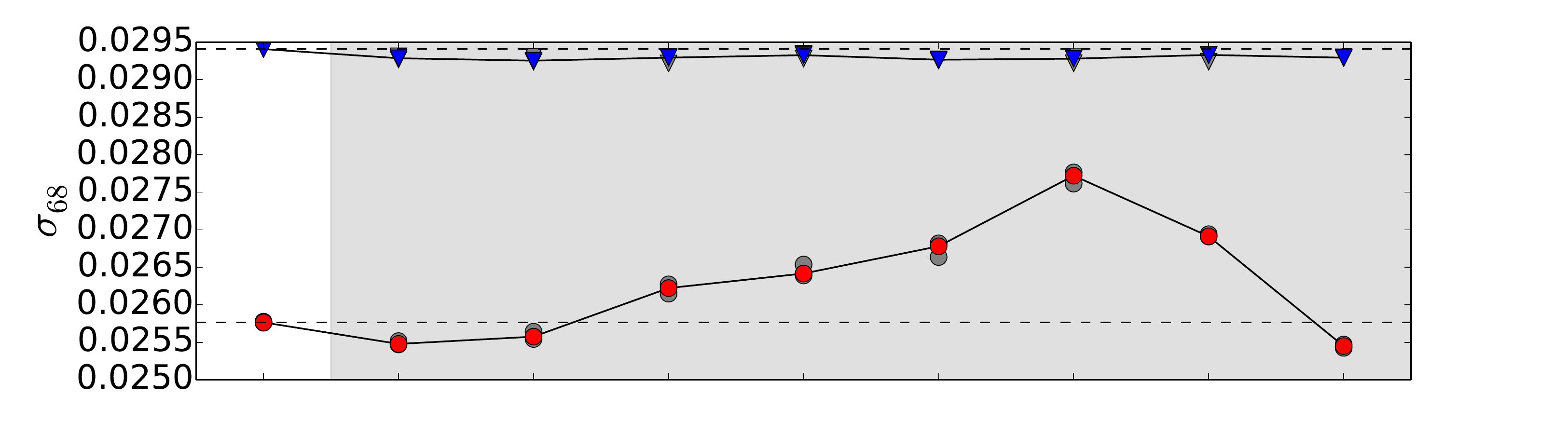}
\includegraphics[scale=0.29, clip=true, trim=8 1 90 16]{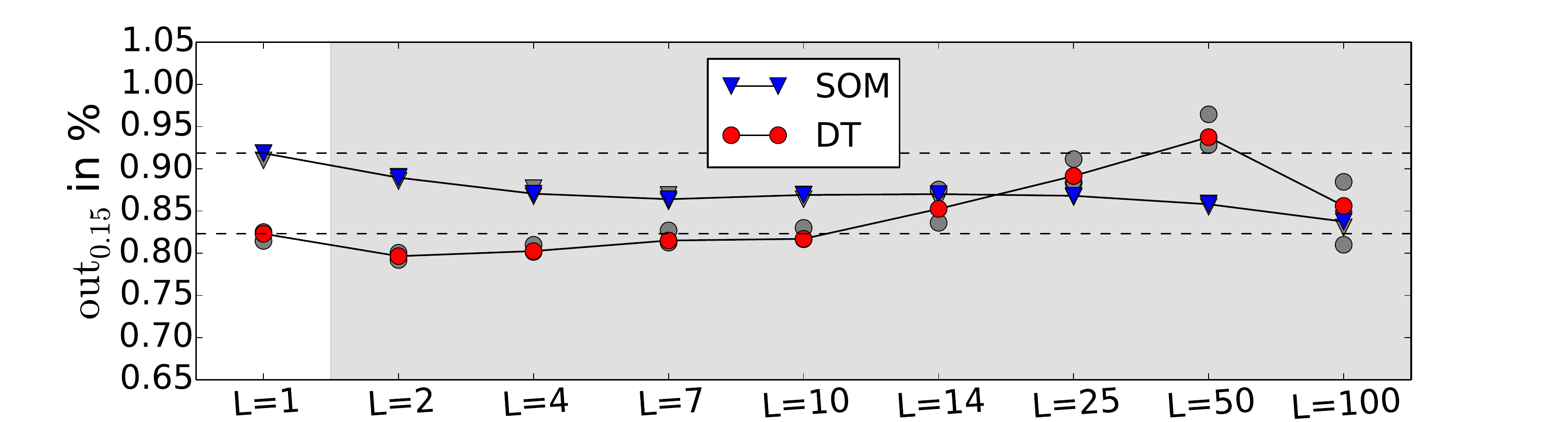}
\caption{ \label{comp_som} The absolute metric values for different stacking configurations which have the following fixed number of base learners $P=L \times N =100$. We show both the stacking systems with SOMs and decision trees as the base learners, and increase the number of layers $L$ and reduce the number of learners per layer $N$ from left to right. We show the metric values as calculated on the validation sample. The dashed horizontal lines mark the performance of using the comparison system without any stacking layers, $L=1$, and the grey regions correspond to algorithms enhanced by extra stacking layers. {\rr The grey data show the spread of the metric values, and the colored points show the median.}}  
\end{figure}

We now concentrate on one of the sets of diagonal lines as presented in Fig. \ref{gridsearchImg}, which correspond to a total number of learners $P=100$, and present the absolute (not relative) metric values for different stacking configurations in Fig. \ref{comp_som}. We show the metric values for stacked systems using decision trees and SOMs on the same figure, and highlight the comparison architecture with no stacking layers $L=1$ by a transparent background, and highlight extra stacking layers using a grey background. The errors on the data points represent the standard deviation of metric values calculated after training three different machines using bootstrap re-sampled training data, and then measuring the metric values on the validation sample. {\rr The grey data show the spread of the metric values, and the colored points show the median.}  

Examining Fig. \ref{comp_som} we find that almost all machine learning decision tree configurations are superior than systems constructed using SOMs as the base learners for all of the metrics explored. This result is not new \citep{2014MNRAS.438.3409C,2015MNRAS.452.4183H} but it is the first time it has been shown for stacked systems. {\bh An obvious reason for this improvement is because the decision trees use the spectroscopic redshift during training, to group similar data into leaf nodes, whereas SOMs do not. {\rr We note that there is variation in the spread of metric values, which is probably due to the small number of trained systems that were used to produce redshifts from which the metric values are measured. We would expect the errors to converge with more trained systems.}


\subsection{Increasing the complexity of the base learners}
We next explore how robust the results of \S\ref{grid_search} are to deeper decision trees, and to SOMs with more cells and neurons, by increasing the complexity of these base learners. We choose to increase the size of the SOM from 768 to 3072 neurons. We do not explore intermediate SOM sizes because of time required to train these systems. Decision trees are much faster to train, and we therefore explore a range of decision tree depths, by modifying the hyper-parameter for the number of objects on each leaf node.

We present the results of the analysis comparing the SOM depths in Fig. \ref{comp_som1}. Each panel shows the metric name on the y-axis label and the absolute (not relative) values of the metric values, and the x-axis shows the increase in the number of stacking layers, and therefore the decrease in the number of learners per layer. We show the value of the comparison metric as the $L=1$ data point and have indicated a horizontal dotted line across the panel from this comparison value in order to help guide the eye. The triangular (square) data points show the SOM base learner with 768 (3072) neurons.  {\rr The grey data show the spread of the metric values from each of the trained machines, and the colored points show the median values.}  

\begin{figure}
\includegraphics[scale=0.29, clip=true, trim=8 14 90 4]{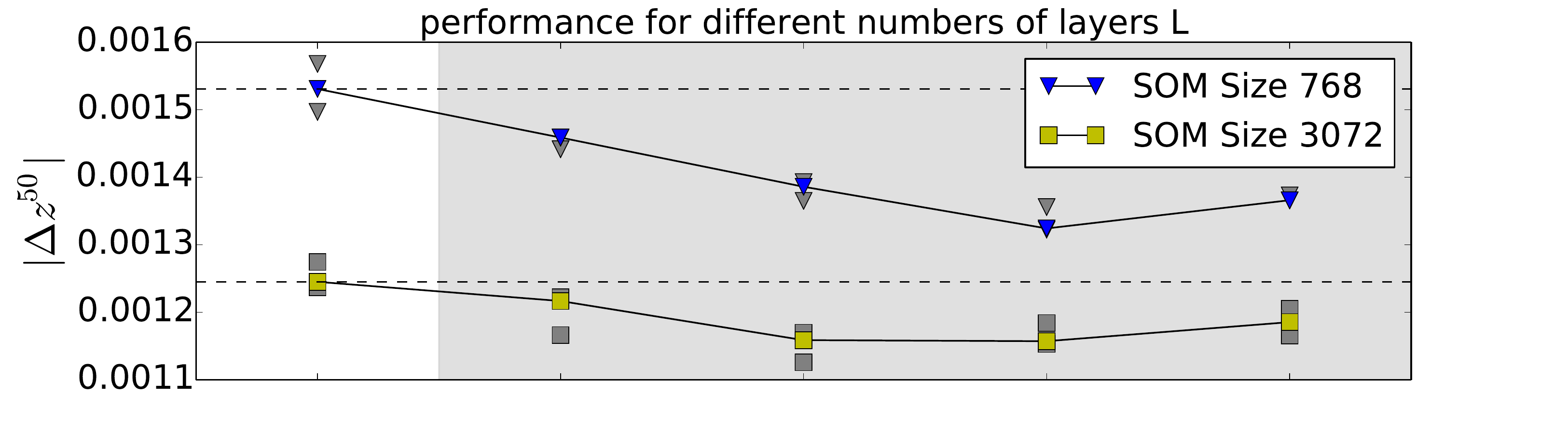}
\includegraphics[scale=0.29, clip=true, trim=8 14 90 16]{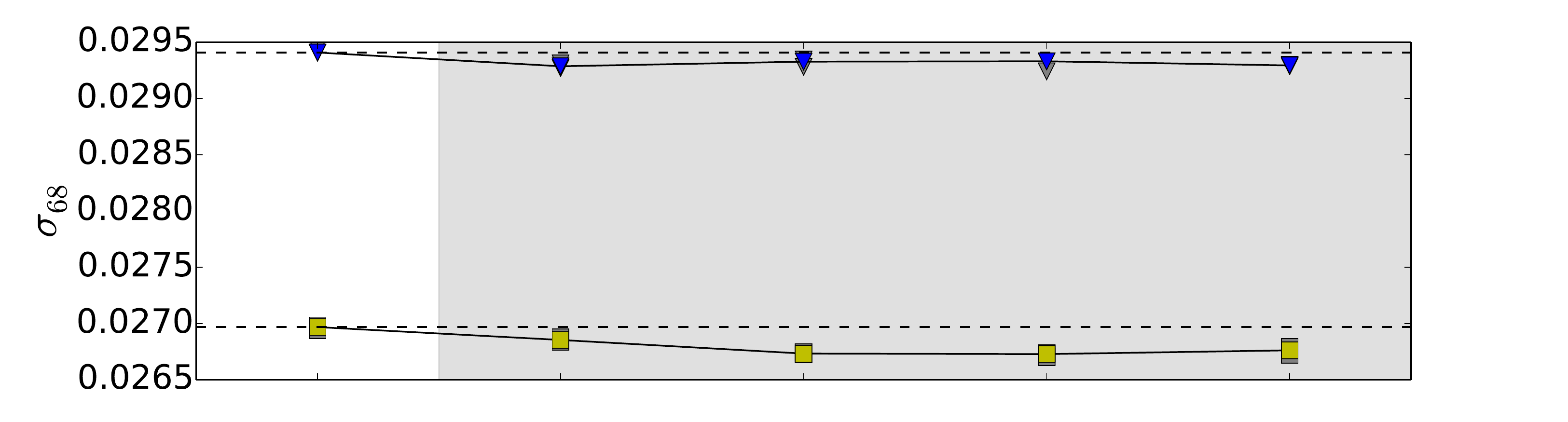}
\includegraphics[scale=0.29, clip=true, trim=8 1 90 16]{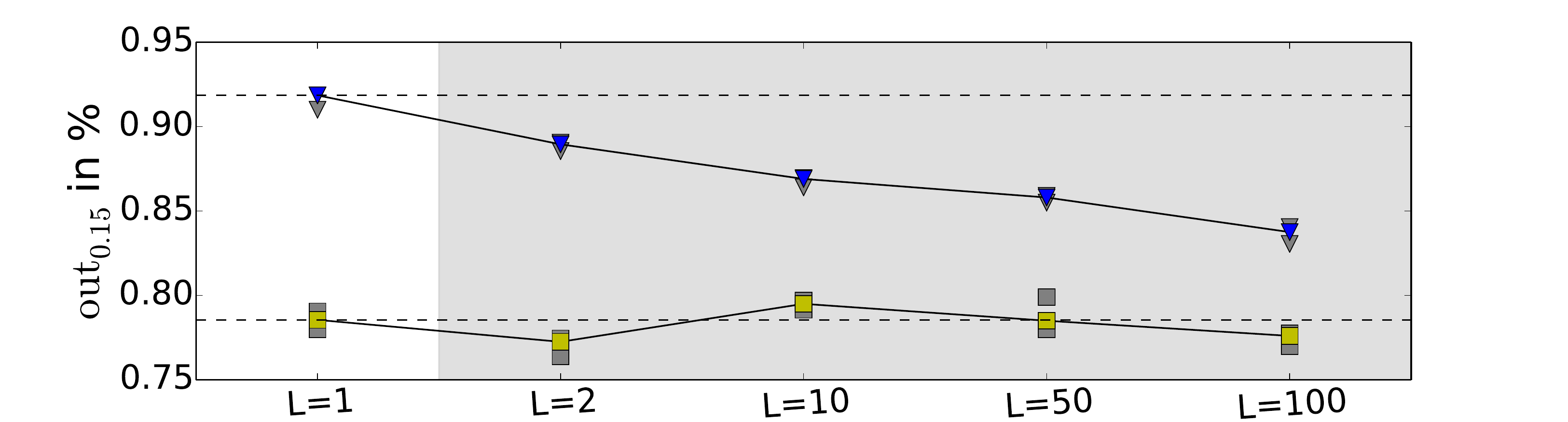}
   \caption{ \label{comp_som1} The effect on the metric values when using a SOM with more neurons as a base learner in the stacking system. The metric names are shown on the y-axis label. We fix the total number of learners $P$ to be 100, and explore a range of values for the number of layers $L$, and the number of learners per layer $N$, such that $P=L \times N = 100$.  {\rr The grey data show the spread of the metric values from each of the trained machines, and the colored points show the median values}}  
\end{figure}

We find that increasing the size of SOM improves all of the metric values by approximately ten percent, irrespective of the choice of stacking architecture. We also find that stacking either improves the metric values by a few percent, or does not degrade the metric values perceptibly. From this analysis, and that presented in \S \ref{grid_search} we conclude that SOMs systems always improve when used in conjunction with stacking, and that the best choice of stacking architectures  for SOMs are either to have one SOM per layer and many layers, or have approximately equal numbers of layers and learners per layer. 

We next increase the complexity of the decision trees, by decreasing the minimum number of objects per final leaf node. In this analysis we choose to explore two stacking architectures with $P=400$, namely $P=L\times N=1\times 400 = 400$ and $P=2\times 200 = 400$, which correspond to having no stacking layers and 400 learners, and one additional stacking layer and 200 learners per layer. 
 
We present the results of the analysis in Fig. \ref{minleaf}, which shows the relative metric values of both stacking architectures as the complexity of the base learner decision trees are increased. The metric values are computed relative to the comparison system which is the system without any additional stacking layers, and with the largest number of objects per leaf node.  {\rr The contours represent the standard deviation from each of the three trained machines.} 
\begin{figure}
\includegraphics[scale=0.32, clip=true, trim=5 24 75 38]{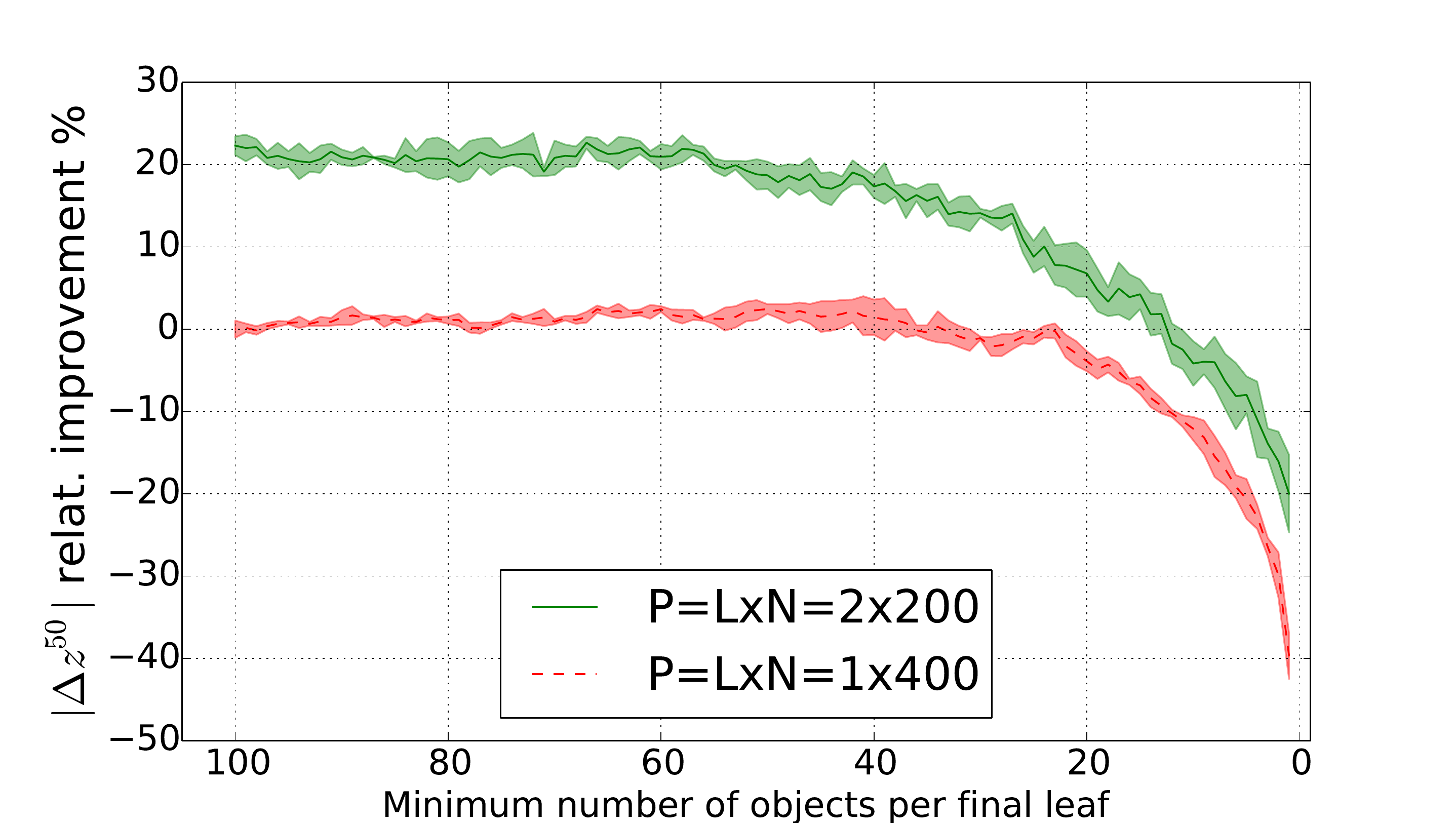}
\includegraphics[scale=0.32, clip=true, trim=5 24 75 38]{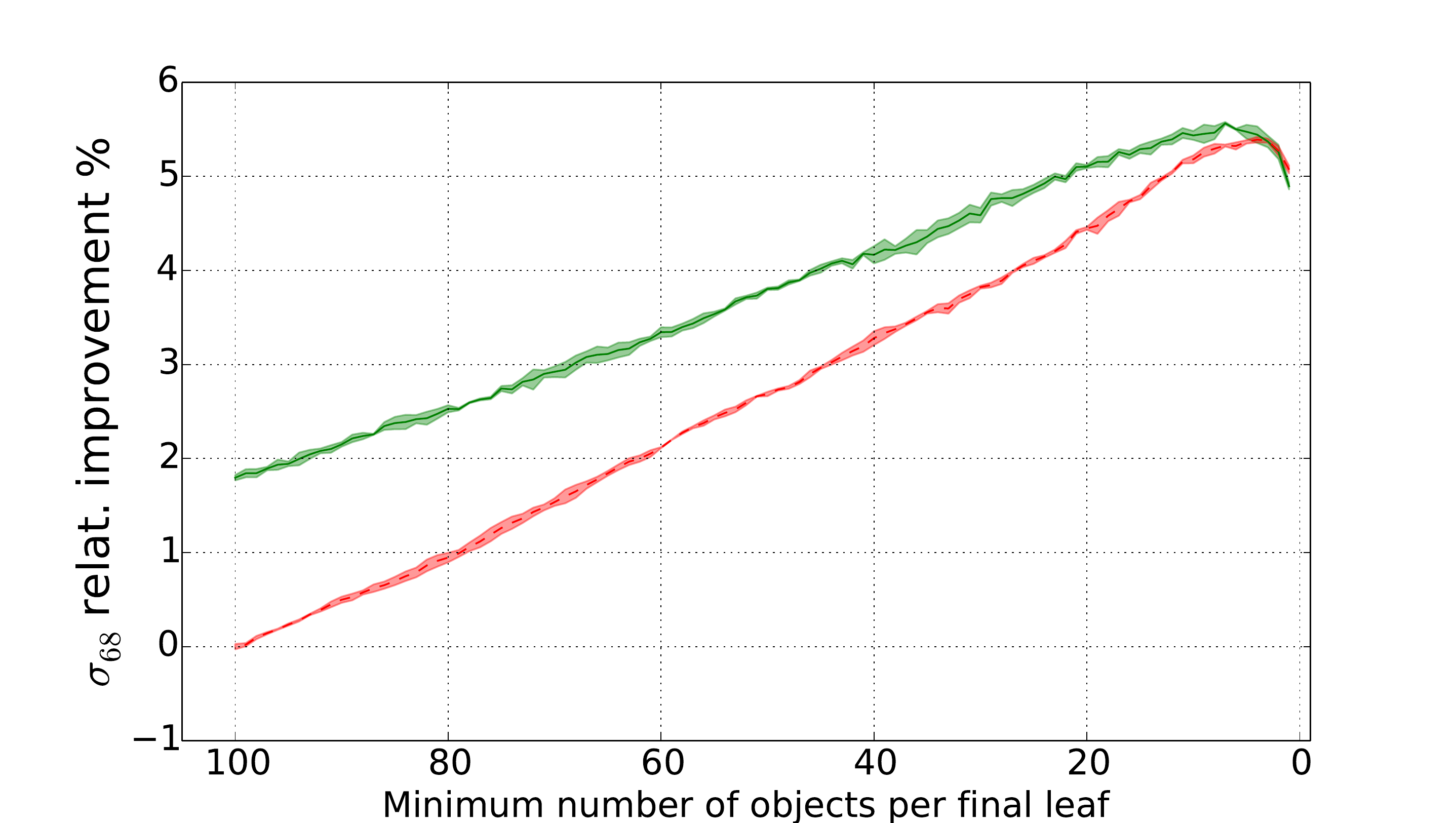}
\includegraphics[scale=0.32, clip=true, trim=5 0 75 38]{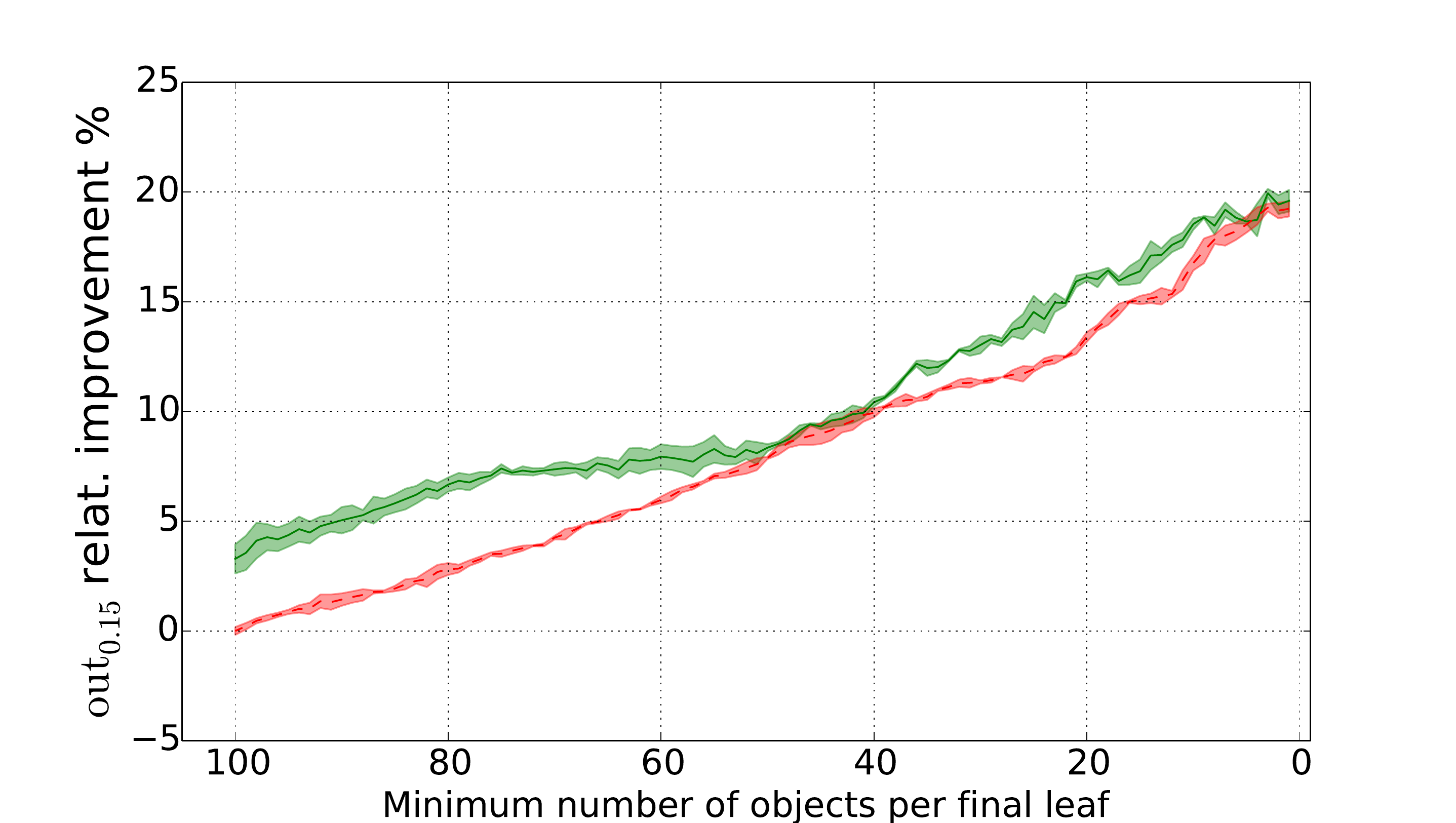}
\caption{ \label{minleaf} The relative metric values of both stacking architectures as the complexity of the decision tree base learners are increased. The metric values are computed relative to the comparison system which is the system without any additional stacking layers, and with the largest number of objects per leaf node. {\rr The contours represent the standard deviation calculated from the three trained machines.} Two stacking architectures are presented with $P=1\times 400 = 400$ and $P=2\times 200 = 400$.
}  
\end{figure}

We again find that stacking always helps to improve the metric values, however when using collections of decision trees as the base learner, we find that the improvement is only a few percent for the metrics $\sigma_{68}$ and the outlier fraction, and up to 20 percent for the median. We note that the results seen in Fig. \ref{minleaf} are also consistent with Fig. \ref{gridsearchImg}. We find that for the metrics  $\sigma_{68}$ and the outlier fraction the systems with more complex decision tree base learners always improves compared to the decision tree with more objects on each leaf node. However this improvement decreases as the decision trees approach purity, i.e., the number of objects on each leaf node approaches 1. This degradation in results is caused by the decision trees over-fitting the training data, and therefore not generalising well to the test data. We find that the median metric is strongly effected by the choice of decision tree complexity, with more complex trees producing much worse estimates of the median value, as compared to both the comparison system and to less complex systems. 

From these analyses we conclude that having a machine learning system with at least one additional layer of stacking improves the redshift estimates when using decision trees as base learners, however the base learners should not {\rr be} too complex.

\subsection{Stacking with strong learners}
\label{comwithad}
In this section we explore how stacking architectures can also improve the redshift measurements of systems which already perform very well, e.g., when using strong learners as the base learners.  We use the AdaBoost algorithm as the base strong learner. Each AdaBoost learner creates an ensemble of  decision trees. In this section the number of base learners $P$ is given by $P=L \times N\times M$, where each layer $L$ consists of $N$ learners, and each learner is a collection of $M$ decision trees. We make comparison systems which are constructed using AdaBoost without any additional stacking layers ($L=1$) but with additional learners, such that the total number of base learners $P$ in each system is the same. We again note that each AdaBoost learner is trained using a bootstrap re-sampled training sample. 

We further fix the depth of the decision trees which are combined with AdaBoost by setting the hyper-parameter corresponding to the minimum number of objects per leaf node to the value of 20. This choice is motivated by examining Fig. \ref{minleaf} for which this value corresponds to a reasonable improvement, or at least minimal degradation, in the values of the chosen metrics, for the collections of decision trees. In what follows we compare stacking with one additional layer ($L=2$) with not performing any stacking ($L=1$). We leave the exploration of deeper than $L=2$ stacking architectures to future work, however we do not expect to see substantial improvements by adding more stacking layers because of the lack of substantial improvement found using collections of decision trees.

For computation cost reasons we set the total number of AdaBoost learners to be $P= [1 $ or $ 2] \times [100 $ or $ 50]  \times M=100 \times M$. We explore the effect of adding more decision trees $M$ to the Adaboost learners, and present the results of the both the stacked and unstacked analysis in Fig. \ref{adab}. We choose to show the relative performance in the metric values with respect to the comparison AdaBoost system which has the least number of decision trees with no additional stacking layers. The metric name is shown on the y-axis label of Fig. \ref{adab} and {\rr the contours represent the standard deviation from each of the three trained machines.}

\begin{figure}
\includegraphics[scale=0.31, clip=true, trim=0 24 75 38]{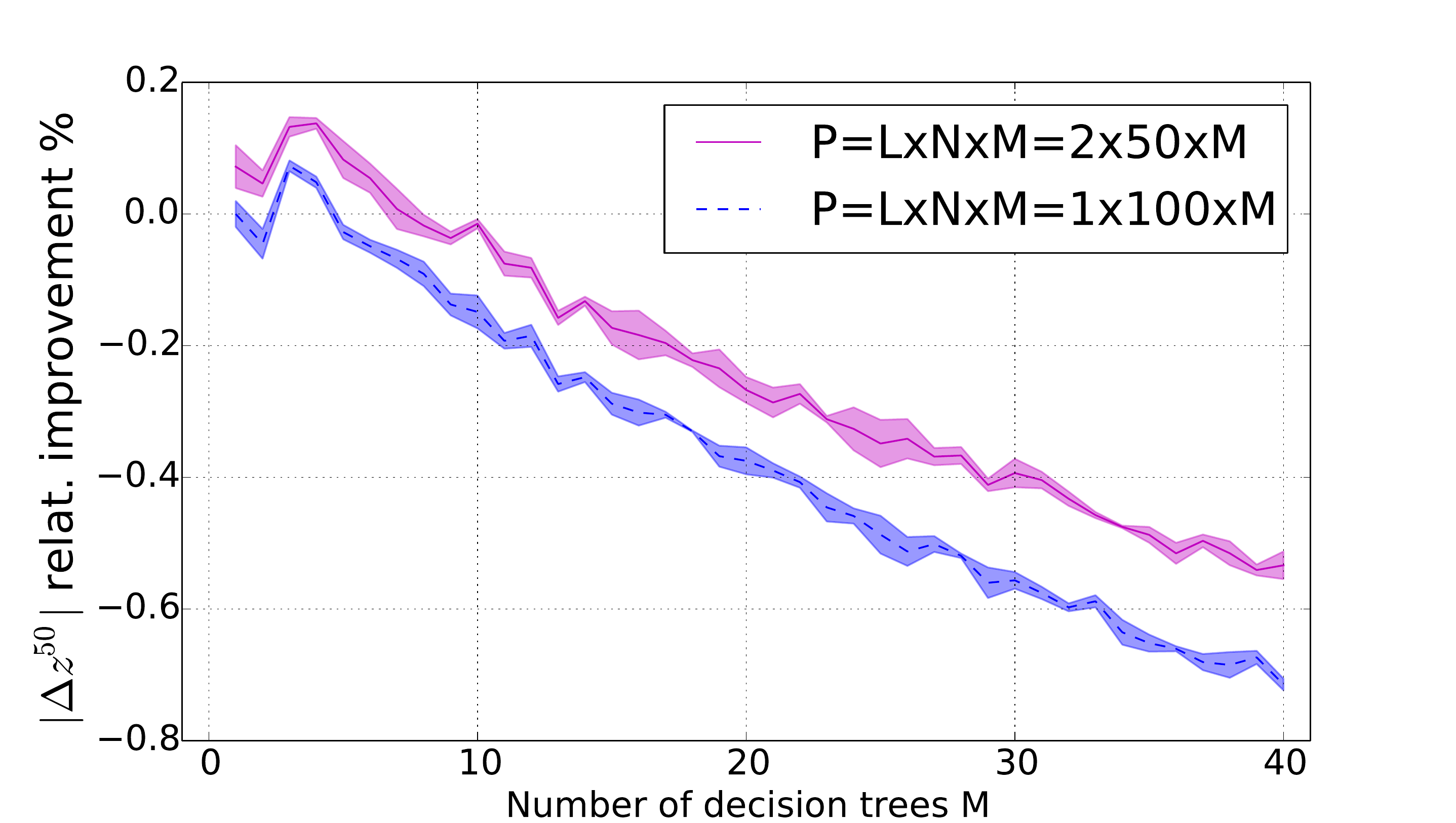}
\includegraphics[scale=0.31, clip=true, trim=0 24 75 38]{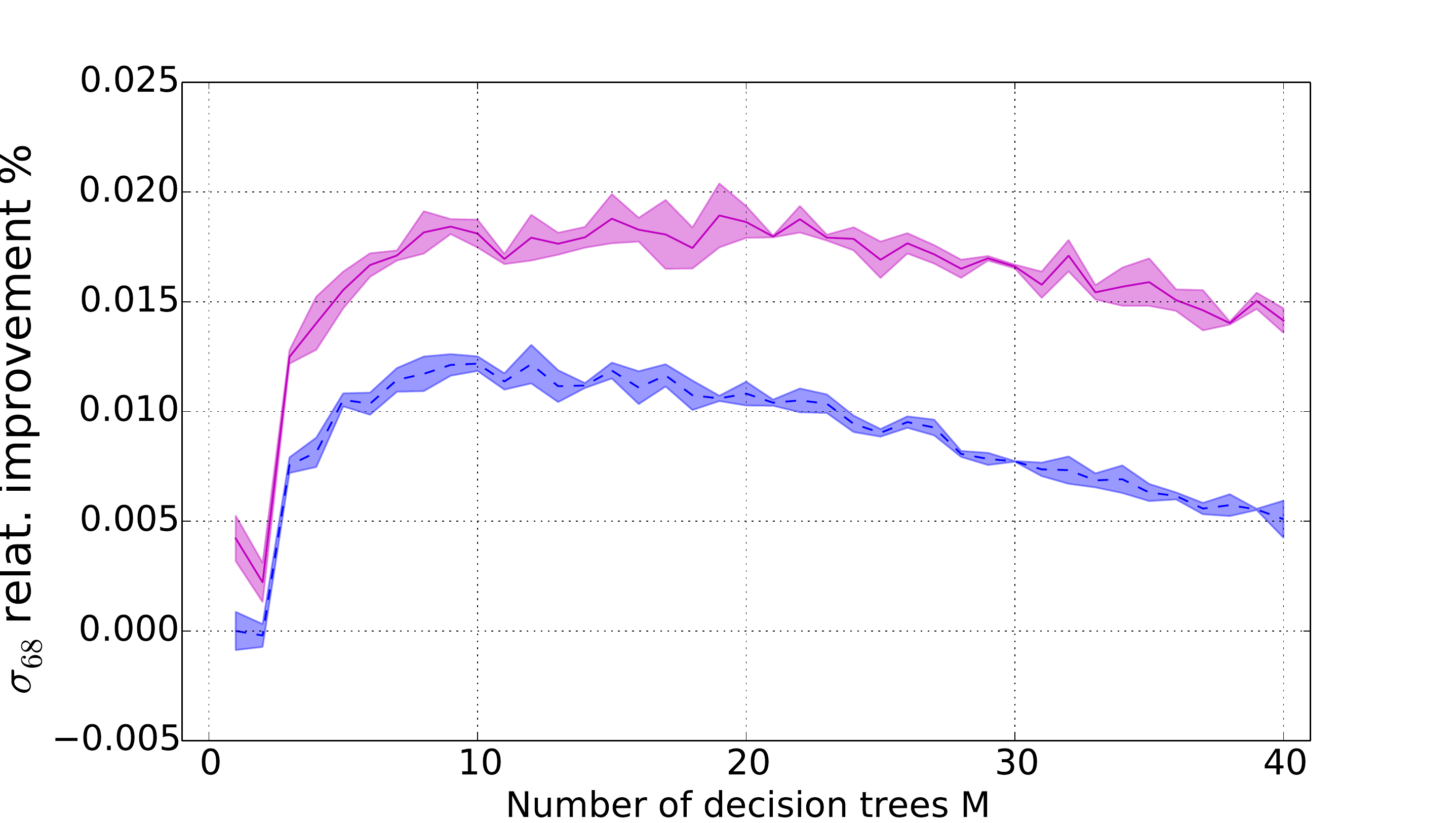}
\includegraphics[scale=0.31, clip=true, trim=0 0 75 38]{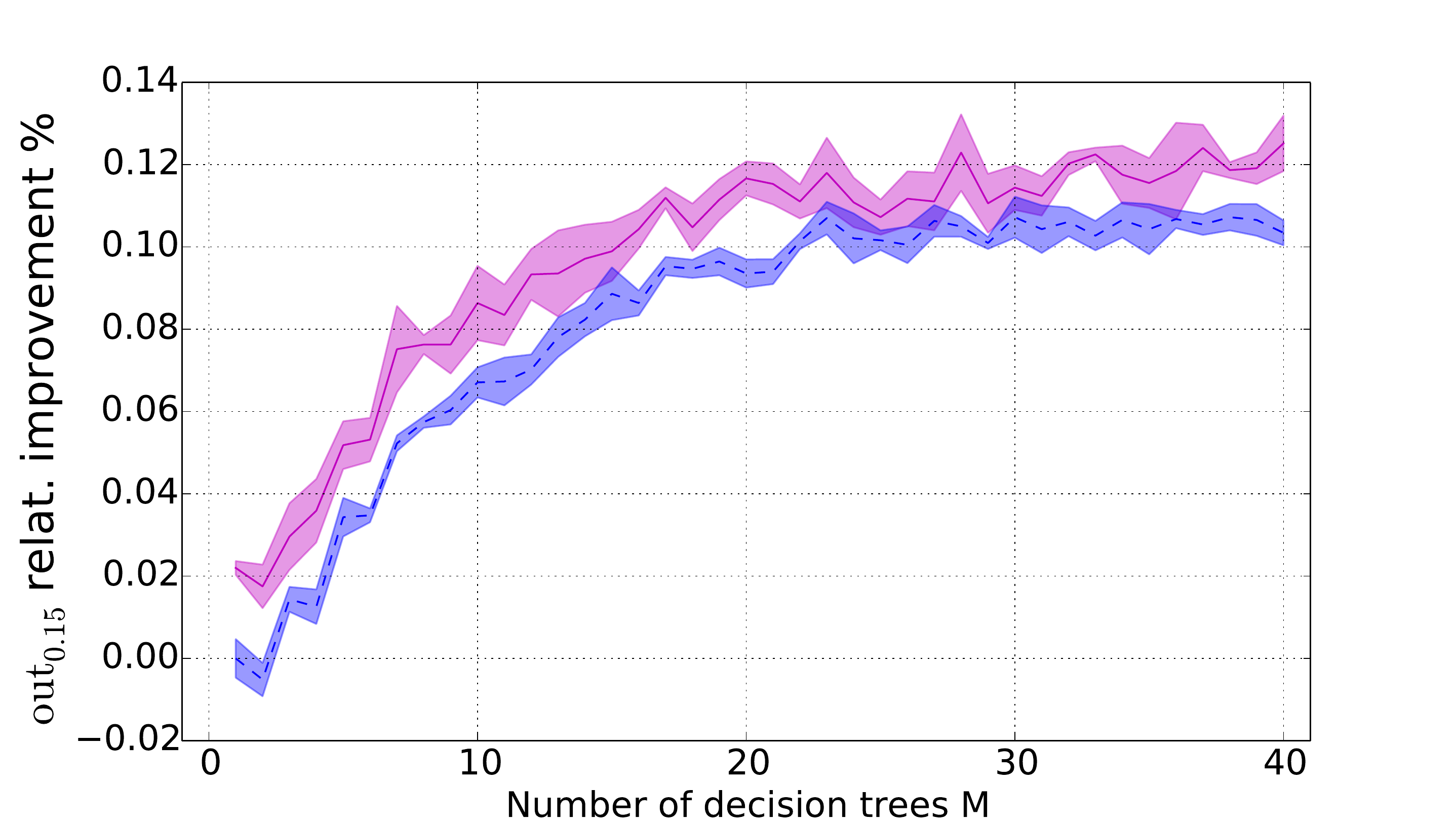}
\caption{ \label{adab} The effect of adding more decision trees $M$ to the Adaboost learners with and without additional layers of stacking. We show the relative performance in the metric values with respect to the comparison AdaBoost system with the least number of decision trees and with no additional stacking layers. {\rr The contours represent the standard deviation from each of the three trained machines.}
} 
\end{figure}

Examining Fig. \ref{adab} we find the following two main results; stacking always improves the metric values compared with not stacking, however this improvement is a modest few percent, and secondly, the stacked AdaBoost algorithm is more accurate when less decision trees $M$ are combined in correspondingly more learners $N$. We also find that there is no obvious architecture that achieves an optimal balance between the number of learners per layer $N$, and number of decision trees per learner $M$. This choice would have to be found, as performed  here, by exploring this parameter space and measuring the chosen metric on a validation sample.

\subsection{Summary of stacking results}
\label{summarytest}
In this section we summarise the results of the previous sections and demonstrate how generalisable these results are. We first select the best machine learning architecture hyper-parameter values, and the best stacking architectures parameters using the validation sample, and present the hyper-parameters of the systems in Table \ref{hyperbestperf1}. We next retrain 10 systems each of which use different bagged re-sampled training data. We then pass the large sample of test data through each of the 10 systems and measure the metric values. We now report and present the mean and standard deviation of these 10 values, as measured on the test {\rr sample in Table \ref{final_results} and Fig. \ref{rankfeat}.} In Table \ref{final_results} we highlight the best metric values in bold, however note that some of the other values are statistically indistinguishable from each other.  We reiterate that the metrics are measured on the test sample which is not used during the machine training or the hyper-parameter selection process.

Fig. \ref{rankfeat} shows how the metric values are all improved when we include stacking layers for each of the machine learning architectures explored in this work. We use the following acryonms for the name of the machine learning architecture on the x-axis label: AdaBoost AD, decision trees DT, and self organising maps SOM, and these algorithms combined with stacking layers are denoted by `+st'. We also further highlight the (un)stacked architectures by the 
(white) grey backgrounds. We connect the metric values between the unstacked and stacked architectures, and further highlight the unstacked values by the dotted lines for each machine learning architecture. We use the y-axis labels to show the different metrics presented in each panel, and we have inflated the error bars by a factor of 5 for viewing purposes.

\begin{figure}
\includegraphics[scale=0.3, clip=true, trim=0 20 85 18]{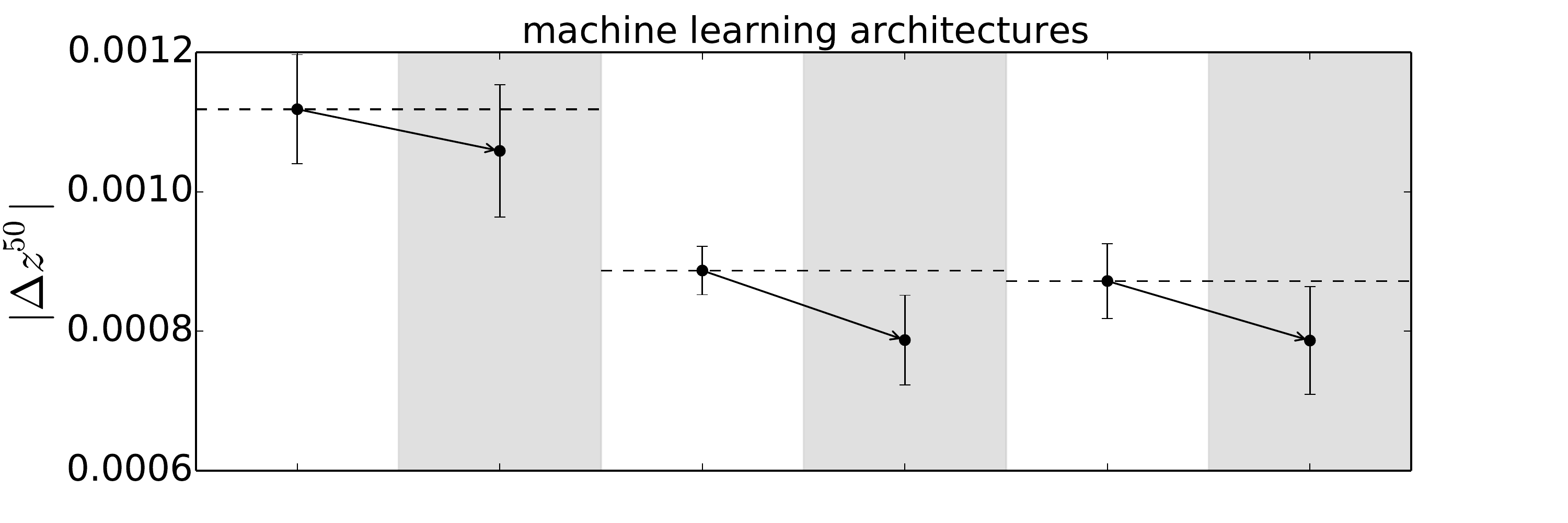}
\includegraphics[scale=0.3, clip=true, trim=0 20 85 18]{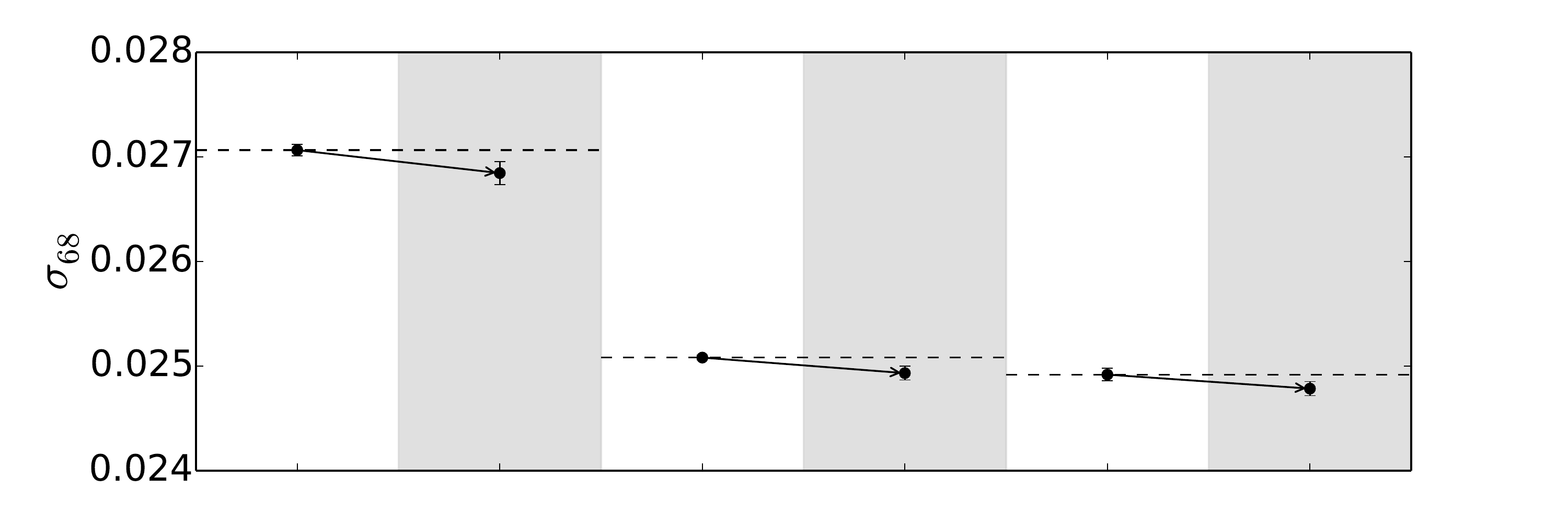}
\includegraphics[scale=0.3, clip=true, trim=0 1 85 18]{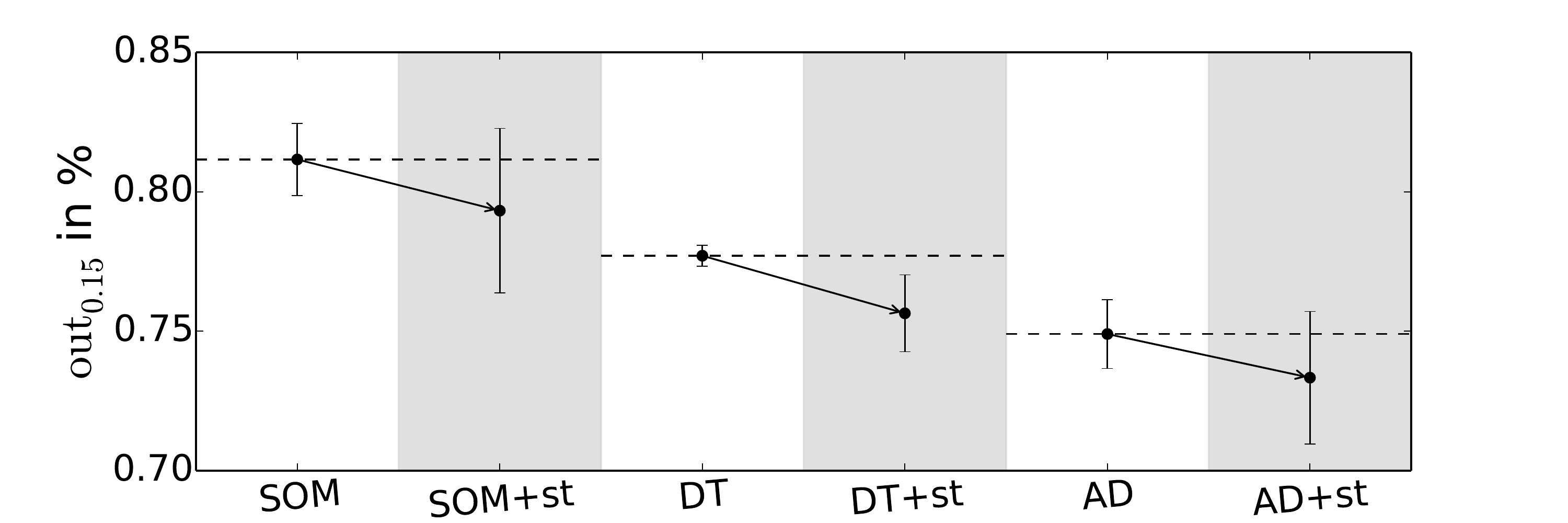}
\caption{ \label{rankfeat} Summary statistics of the improvement for each of the machine learning algorithms when include additional stacking layers. We use the following acryonms for the name of the machine learning algorithm on the x-axis label: AdaBoost AD, decision trees DT, and self organising maps SOM, and these algorithm combined with stacking layers are shown by +st. We also further highlight the (un)stacked architectures by the 
(white) grey backgrounds. The metric values are connected between the unstacked and stacked architectures, and the unstacked values are further highlighted by the dotted lines for each machine learning architecture. We use the y-axis labels to show the different metrics presented in each panel.
The error bars have been inflated by a factor of 5 for viewing purposes.}  
\end{figure}

Fig. \ref{rankfeat} clearly shows that there is always an improvement available, to each of the machine learning algorithm explored in this work, if some of the learners first produce redshift estimates and then those estimates are passed as additional input features into an additional layer of learners. We have shown that stacking architectures are able to extract further improvements in the explored metric values, {\rr with only a very modest increase in the additional computational requirements to the user.}

\begin{table}
\centering
\renewcommand{\footnoterule}{}
\begin{tabular}{ll}
Algorithm & Configuration\\
\hline
\hline
SOM without stacking & $N_{n}= 3072$  \\
\ & $P=100$, $L=1, N=100$ \\
\hline
SOM with stacking & $N_{n}= 3072$ \\
\ & $P=100$, $L=100, N=1$ \\
\hline
Decision trees without stacking&$N_{leaf}=20$\\
\ & $P=400, L=1, N=400$ \\
\hline
Decision trees with stacking&$N_{leaf}=20$\\
\ & $P=400$, $L=2, N=200$ \\
\hline
Adaboost without stacking  &$N_{leaf}=20$\\
\ &  \\
\ & $P=100, L=1, N=20, M=5$ \\
\hline
AdaBoost with stacking &$N_{leaf}=20$ \\
\ & $P=100$, $L=2, N=10, M=5$ \\
\end{tabular}
\caption{Hyper-parameter values for the best-performing algorithms listed in Table \ref{final_results}. We denote the number of neurons of the SOM by $N_n$, the number of training examples per leaf node of a decision tree by $N_{leaf}$, the number of stacking layers $L$, the number of learners per layer $N$ and the number of decision trees per AdaBoost base learner $M$. The total number of base learners is given by $P=L \times N (\times M)$.}
\label{hyperbestperf1}
\end{table}

\begin{table}
\centering
\renewcommand{\footnoterule}{}
\begin{tabular}{l c c c }
\hline
Algorithm & $|\Delta z^{50}|$ & $\sigma_{68}$  & ${\rm out}_{0.15} \%$  \\
 \hline  \hline 
SOM without stacking & 0.0011 & 0.0270 & 0.811  \\
SOM with stacking & 0.0011 & 0.0268 & 0.793 \\
Decision trees without stacking  & 0.0009 & 0.0251 & 0.777 \\
Decision trees with stacking & \textbf{0.0008} & 0.0249 & 0.756 \\
AdaBoost without stacking & 0.0009 & 0.0249 & 0.749  \\
AdaBoost with stacking & \textbf{0.0008} & \textbf{0.0248} & \textbf{0.733} \\
 \hline 
\end{tabular}
\caption{A summary of the performance results as calculated on the independent test sample. We present the mean of the metric values using 10 different machines trained with fixed hyper-parameters (see Table \ref{hyperbestperf1}). The bold values represent the best results per column.}
\label{final_results}
\end{table}
\section{discussion}
\label{discussion}
{\bh We have shown that stacking is an effective way to improve machine learning predictions for photometric redshift analysis applied to SDSS data. We now present a discussion of why this may be the case.

If we concentrate on the middle panel of Fig. \ref{adab} which shows the dispersion $\sigma_{68}$, of the redshift scaled residuals as a function of the number of training data per leaf node, which describes the depth of the tree and is a measure for the complexity of the system. This panel shows the relative improvement in $\sigma_{68}$, and therefore larger y-values correspond to smaller values of $\sigma_{68}$. The dispersion is an estimate of the standard deviation of the distribution, which is related to the mean squared error (MSE). This enables us to view this panel as (the y-axis inverse of) a MSE against model (or tree) complexity diagram. This is then nothing more than the usual learning curve from machine learning.

If we concentrate first on the system without any additional stacking layers, we see that for the low complexity model, in the left of the panel, there is a high bias: meaning the model does not fit the data very well. In the right of the panel, the model fits the data point very well, so we are in the low bias regime, and potentially high-variance: meaning over fitting of the data. However because we are using collection of decision trees, we do not find that the system strongly over fits, and therefore the variance is always low. Over fitting would occur, for example, if we were to only use one decision tree. Therefore the difference between the left and the right hand sides show how the increased complexity of the model decreases the value of bias. 

If we next concentrate on the left of the panel and examine the difference between the stacked $L=2$, and unstacked $L=1$, systems, we find that the more complex system, i.e. that which has two learning rounds, has a larger improvement in the value of $\sigma_{68}$, and therefore a lower MSE. This improvement must be due to the increased complexity leading to a faster reduction in bias. If we now concentrate on the right side of the panel, we see that both systems converge to the same values of MSE, and therefore they now have the same, very low, value of bias. 

We therefore conclude, with a somewhat obvious statement, that stacking increases the complexity of the model, and therefore can provide a better fit to the data. 
}
\section{conclusions}
\label{conclusions}
We have introduced `stacking' architectures for the task of photometric redshift estimation using machine learning, and have shown the benefit of using such stacked systems on common metrics. Stacking is a generic method to use the machine learning predictions obtained from one training round as an additional input into a different training round, or training `layer'. Any machine learning algorithm can be used as the base learner which are then further combined in each of the stacking layers. For a schematic diagram of the stacking architecture used in this paper see Fig. \ref{flowchart}.

We explore the use of various stacking architecture configurations, such as the number of stacking layers, and the method to combine and pass predictions across different layers, and we find that all explored machine learning algorithms improve by having at least one additional stacking layer for the task of photometric redshift estimation. The improvement is modest and varies between 2 and 20 percent depending on the measured metric and stacked algorithm.

We apply stacking to three very different machine learning algorithms, two are based on decision trees in a supervised learning setting, and the third uses self-organising maps (SOMs) in an unsupervised learning setting. We further split these algorithms into so called `weak learners' which are models with low predictive power, such as individual decision trees or a  SOM, and `strong learners', for example using the AdaBoost algorithm to combine decision trees. The AdaBoost algorithm combines sequentially constructed decision trees, with each tree concentrating on the most difficult training examples from the output of the previous tree.

We begin by exploring a grid of different stacking architectures. We use an independent validation sample to calculate the metrics values obtained by each stacked architecture, and to identify winning architectures. We find that some algorithms, such as SOMs perform better by stacking many layers with a low number of learners per layer, and other algorithms, such as decision trees, perform better by having many learners per layer and only a few layers.

We choose to quantify the machine learning redshift $z_{ML}$, prediction performance 
by constructing the redshift scaled residual distribution, defined as $\Delta_z = (z-z_{ML})/(1+z)$, and measuring the following three metrics; the median, 68\% dispersion, and outlier fraction, defined as the fraction of data with $|\Delta_z|>0.15$. We note that the 68\% dispersion $\sigma_{68}$ is equivalent to the standard deviation for a Gaussian distribution. Depending on the science case, one may instead wish to measure the redshift prediction ability of the recovered full redshift distribution of a test sample. We leave the exploration of other metrics such as this to future work.

In this work we have explored only a few of the tunable hyper-parameters of the base learning algorithms, such as the depth of the trees, and the number of neurons in each SOM. We have concentrated instead on the hyper-parameters of the stacking system itself, such as the number of stacking layers, the number of base learners per layer, and the method that combines the predictions and passes them from one layer to the next. There are many other algorithm hyper-parameters that could be explored in future work, however given the analysis presented here, we would still expect stacking to improve the predictions, even for a richer hyper-parameter space. One could also explore the combination of different sets of algorithms in the same layer, or change the base learner algorithm between layers. In this work we have explored three very different machine learning algorithms, and therefore expect that other algorithms will also likely shown an improvement if used in a stacked system. We have also shown that for a fixed number of learners, there is always an improvement in the measured metrics when introducing at least some amount of stacking. 

We have only explored stacking for the machine learning task of photometric redshift prediction. One could equally explore how stacking effects other common machine learning tasks, such as star and galaxy separation \citep[e.g.,][]{1997daa..conf...43L,2009arXiv0910.3770Y}, or optimal target selection \citep[e.g.,][]{2015arXiv150806280H}.

We are currently applying these techniques to other datasets, and machine learning tasks to determine if stacking is a generic method to further improve the performance of a set of machine learning algorithms. As with all machine learning analysis, it is highly recommend to explore stacking for each problem, and for many datasets before assuming that improvement is always achieved. However this work shows that stacking has the potential to further improve the predictions obtained by even very strong base learners.



\section*{Acknowledgments} 
\label{ack}
Funding
for the SDSS and SDSS-II has been provided by the Alfred
P. Sloan Foundation, the Participating Institutions, the
National Science Foundation, the U.S. Department of
Energy, the National Aeronautics and Space Administration,
the Japanese Monbukagakusho, the Max Planck
Society, and the Higher Education Funding Council for
England. The SDSS Web Site is http://www.sdss.org/.

\appendix
\section{} 
\label{appendix}
\subsection{MYSQL QUERIES}
\label{mysql}
We run the following {\tt MySQL} query in the SDSS Data Release 10 scheme to extract observational data used throughout this project. We further post process this data as described in \S\ref{featscal}.

\begin{verbatim}
SELECT s.specObjID, s.objid, s.ra,s.dec, 
s.z as specz, s.zerr as specz_err,
s.type as specType, q.type as photpType,
q.extinction_u,q.extinction_g,q.extinction_r,
q.extinction_i,q.extinction_z,
q.psfMag_u,q.psfMagErr_u, 
q.psfMag_g,q.psfMagErr_g, 
q.psfMag_r,q.psfMagErr_r, 
q.psfMag_i,q.psfMagErr_i, 
q.psfMag_z,q.psfMagErr_z, 
    
INTO mydb.specPhotoDR10v4 FROM SpecPhotoAll AS s 
   
JOIN photoObjAll AS q 
ON s.objid=q.objid  AND q.cModelMag_g>0 
AND q.cModelMag_r>0 AND q.cModelMag_z>0 

LEFT OUTER JOIN Photoz AS p ON s.objid=p.objid
\end{verbatim}

\bibliographystyle{mn2e}
\bibliography{photoz}

\begin{thebibliography}{41}
\expandafter\ifx\csname natexlab\endcsname\relax\def\natexlab#1{#1}\fi

\bibitem[{{Ahn} {et~al}\mbox{.}(2014){Ahn}, {Alexandroff}, {Allende Prieto},
  {Anders}, {Anderson}, {Anderton}, {Andrews}, {Aubourg}, {Bailey}, {Bastien},
  \& et~al.}]{2014ApJS..211...17A}
{Ahn} C.~P. {et~al.}, 2014, \apjs, 211, 17

\bibitem[{{Bonnett}(2015)}]{2013arXiv1312.1287B}
{Bonnett} C., 2015, \mnras, 449, 1043

\bibitem[{{Bonnett} {et~al}\mbox{.}(2015){Bonnett}, {Troxel}, {Hartley},
  {Amara}, {Leistedt}, {Becker}, {et~al.}}]{2015arXiv150705909B}
{Bonnett} C., {Troxel} M.~A., {Hartley} W., {Amara} A., {Leistedt} B., {Becker}
  M.~R., {et~al.}, 2015, ArXiv:1507.05909

\bibitem[{Breiman(2001)}]{RandoMforests}
Breiman L., 2001, Machine Learning, 45, 5

\bibitem[{Breiman {et~al}\mbox{.}(1984)Breiman, Friedman, Olshen, \&
  Stone}]{ig}
Breiman L., Friedman J.~H., Olshen R.~A., Stone C.~J., 1984, Classification and
  Regression Trees, no, ed. Wadsworth International Group, Belmont, CA

\bibitem[{{Brescia} {et~al}\mbox{.}(2014){Brescia}, {Cavuoti}, {Longo}, \& {De
  Stefano}}]{2014A&A...568A.126B}
{Brescia} M., {Cavuoti} S., {Longo} G., {De Stefano} V., 2014, \aap, 568, A126

\bibitem[{{Carliles} {et~al}\mbox{.}(2008){Carliles}, {Budav{\'a}ri}, {Heinis},
  {Priebe}, \& {Szalay}}]{2008ASPC..394..521C}
{Carliles} S., {Budav{\'a}ri} T., {Heinis} S., {Priebe} C., {Szalay} A., 2008,
  in Astronomical Society of the Pacific Conference Series, Vol. 394,
  Astronomical Data Analysis Software and Systems XVII, {Argyle} R.~W.,
  {Bunclark} P.~S., {Lewis} J.~R., eds., p. 521

\bibitem[{{Carrasco Kind} \& {Brunner}(2013)}]{tpz}
{Carrasco Kind} M., {Brunner} R.~J., 2013, \mnras, 432, 1483

\bibitem[{{Carrasco Kind} \&
  {Brunner}(2014{\natexlab{a}})}]{2014MNRAS.442.3380C}
{Carrasco Kind} M., {Brunner} R.~J., 2014{\natexlab{a}}, \mnras, 442, 3380

\bibitem[{{Carrasco Kind} \&
  {Brunner}(2014{\natexlab{b}})}]{2014MNRAS.438.3409C}
{Carrasco Kind} M., {Brunner} R.~J., 2014{\natexlab{b}}, \mnras, 438, 3409

\bibitem[{{Collister} \& {Lahav}(2004)}]{2004PASP..116..345C}
{Collister} A.~A., {Lahav} O., 2004, \pasp, 116, 345

\bibitem[{{Csabai} {et~al}\mbox{.}(2007){Csabai}, {Dobos}, {Trencs{\'e}ni},
  {Herczegh}, {J{\'o}zsa}, {Purger}, {Budav{\'a}ri}, \&
  {Szalay}}]{2007AN....328..852C}
{Csabai} I., {Dobos} L., {Trencs{\'e}ni} M., {Herczegh} G., {J{\'o}zsa} P.,
  {Purger} N., {Budav{\'a}ri} T., {Szalay} A.~S., 2007, Astronomische
  Nachrichten, 328, 852

\bibitem[{{Dieleman}, {Willett} \& {Dambre}(2015){Dieleman}, {Willett}, \&
  {Dambre}}]{2015arXiv150307077D}
{Dieleman} S., {Willett} K.~W., {Dambre} J., 2015, \mnras, 450, 1441

\bibitem[{Dietterich(2000)}]{Dietterich:2000:EMM:648054.743935}
Dietterich T.~G., 2000, in Proceedings of the First International Workshop on
  Multiple Classifier Systems, MCS '00, Springer-Verlag, London, UK, UK, pp.
  1--15

\bibitem[{Drucker(1997)}]{Drucker:1997:IRU:645526.657132}
Drucker H., 1997, in Proceedings of the Fourteenth International Conference on
  Machine Learning, ICML '97, Morgan Kaufmann Publishers Inc., San Francisco,
  CA, USA, pp. 107--115

\bibitem[{{Eisenstein} \& et~al.(2011)}]{2011AJ....142...72E}
{Eisenstein} D.~J., et~al., 2011, \aj, 142, 72

\bibitem[{Freund \& Schapire(1997)}]{Freund1997119}
Freund Y., Schapire R.~E., 1997, Journal of Computer and System Sciences, 55,
  119

\bibitem[{Friedman(2001)}]{friedman2001}
Friedman J.~H., 2001, Ann. Statist., 29, 1189

\bibitem[{{Gerdes} {et~al}\mbox{.}(2010){Gerdes}, {Sypniewski}, {McKay}, {Hao},
  {Weis}, {Wechsler}, \& {Busha}}]{2010ApJ...715..823G}
{Gerdes} D.~W., {Sypniewski} A.~J., {McKay} T.~A., {Hao} J., {Weis} M.~R.,
  {Wechsler} R.~H., {Busha} M.~T., 2010, \apj, 715, 823

\bibitem[{{Gunn} {et~al}\mbox{.}(2006){Gunn}, {Siegmund}, {Mannery}, {Owen},
  {Hull}, {Leger}, {Carey}, {Knapp}, {York}, {Boroski}, {Kent}, {Lupton},
  {Rockosi}, {et~al.}}]{Gunn:2006tw}
{Gunn} J.~E. {et~al.}, 2006, \aj, 131, 2332

\bibitem[{{Hogan}, {Fairbairn} \& {Seeburn}(2015){Hogan}, {Fairbairn}, \&
  {Seeburn}}]{2015MNRAS.449.2040H}
{Hogan} R., {Fairbairn} M., {Seeburn} N., 2015, \mnras, 449, 2040

\bibitem[{{Hoyle}(2016)}]{2015arXiv150407255H}
{Hoyle} B., 2016, Astronomy and Computing, 16, 34

\bibitem[{{Hoyle} {et~al}\mbox{.}(2015{\natexlab{a}}){Hoyle}, {Paech}, {Rau},
  {Seitz}, \& {Weller}}]{2015arXiv150806280H}
{Hoyle} B., {Paech} K., {Rau} M.~M., {Seitz} S., {Weller} J.,
  2015{\natexlab{a}}, ArXiv:1508.06280

\bibitem[{{Hoyle} {et~al}\mbox{.}(2015{\natexlab{b}}){Hoyle}, {Rau}, {Bonnett},
  {Seitz}, \& {Weller}}]{2015arXiv150106759H}
{Hoyle} B., {Rau} M.~M., {Bonnett} C., {Seitz} S., {Weller} J.,
  2015{\natexlab{b}}, \mnras, 450, 305

\bibitem[{{Hoyle} {et~al}\mbox{.}(2015{\natexlab{c}}){Hoyle}, {Rau}, {Paech},
  {Bonnett}, {Seitz}, \& {Weller}}]{2015MNRAS.452.4183H}
{Hoyle} B., {Rau} M.~M., {Paech} K., {Bonnett} C., {Seitz} S., {Weller} J.,
  2015{\natexlab{c}}, \mnras, 452, 4183

\bibitem[{{Hoyle} {et~al}\mbox{.}(2015{\natexlab{d}}){Hoyle}, {Rau}, {Zitlau},
  {Seitz}, \& {Weller}}]{2014arXiv1410.4696H}
{Hoyle} B., {Rau} M.~M., {Zitlau} R., {Seitz} S., {Weller} J.,
  2015{\natexlab{d}}, \mnras, 449, 1275

\bibitem[{{Kim}, {Brunner} \& {Carrasco Kind}(2015){Kim}, {Brunner}, \&
  {Carrasco Kind}}]{2015MNRAS.453..507K}
{Kim} E.~J., {Brunner} R.~J., {Carrasco Kind} M., 2015, \mnras, 453, 507

\bibitem[{Kohonen, Schroeder \& Huang(2001)Kohonen, Schroeder, \&
  Huang}]{Kohonen:2001:SM:558021}
Kohonen T., Schroeder M.~R., Huang T.~S., eds., 2001, Self-Organizing Maps, 3rd
  edn. Springer-Verlag New York, Inc., Secaucus, NJ, USA

\bibitem[{Krizhevsky, Sutskever \& Hinton(2012)Krizhevsky, Sutskever, \&
  Hinton}]{NIPS2012_4824}
Krizhevsky A., Sutskever I., Hinton G.~E., 2012, in Advances in Neural
  Information Processing Systems 25, Pereira F., Burges C., Bottou L.,
  Weinberger K., eds., Curran Associates, Inc., pp. 1097--1105

\bibitem[{{Lahav}(1997)}]{1997daa..conf...43L}
{Lahav} O., 1997, in Data Analysis in Astronomy, {Di Gesu} V., {Duff} M.~J.~B.,
  {Heck} A., {Maccarone} M.~C., {Scarsi} L., {Zimmerman} H.~U., eds., pp.
  43--51

\bibitem[{Lecun \& Bengio(1995)}]{lecun95convolutional}
Lecun Y., Bengio Y., 1995, Convolutional Networks for Images, Speech and Time
  Series, The MIT Press, pp. 255--258

\bibitem[{Li \& Thakar(2008)}]{10.1109/MCSE.2008.6}
Li N., Thakar A.~R., 2008, Computing in Science and Engineering, 10, 18

\bibitem[{{Masters} {et~al}\mbox{.}(2015){Masters}, {Capak}, {Stern}, {Ilbert},
  {et~al.}}]{2015ApJ...813...53M}
{Masters} D., {Capak} P., {Stern} D., {Ilbert} O., {et~al.}, 2015, \apj, 813,
  53

\bibitem[{Pedregosa {et~al}\mbox{.}(2011)Pedregosa {et~al.}}]{scikit-learn}
Pedregosa F., {et~al.}, 2011, Journal of Machine Learning Research, 12, 2825

\bibitem[{{Rau} {et~al}\mbox{.}(2015){Rau}, {Seitz}, {Brimioulle}, {Frank},
  {Friedrich}, {Gruen}, \& {Hoyle}}]{RauEtAllinPrep}
{Rau} M.~M., {Seitz} S., {Brimioulle} F., {Frank} E., {Friedrich} O., {Gruen}
  D., {Hoyle} B., 2015, \mnras, 452, 3710

\bibitem[{{S{\'a}nchez}, {Carrasco Kind} {et~al}\mbox{.}(2014){S{\'a}nchez},
  {Carrasco Kind}, {et~al.}}]{2014MNRAS.445.1482S}
{S{\'a}nchez} C., {Carrasco Kind} M., {et~al.}, 2014, \mnras, 445, 1482

\bibitem[{Smith {et~al}\mbox{.}(2002)Smith {et~al.}}]{Smith:2002pca}
Smith J.~A., {et~al.}, 2002, \aj, 123, 2121

\bibitem[{{Tagliaferri} {et~al}\mbox{.}(2003){Tagliaferri}, {Longo}, {Andreon},
  {Capozziello}, {Donalek}, \& {Giordano}}]{2003LNCS.2859..226T}
{Tagliaferri} R., {Longo} G., {Andreon} S., {Capozziello} S., {Donalek} C.,
  {Giordano} G., 2003, Lecture Notes in Computer Science, 2859, 226

\bibitem[{{Vanzella} {et~al}\mbox{.}(2004){Vanzella}, {Cristiani}, {Fontana},
  {Nonino}, {Arnouts}, {Giallongo}, {Grazian}, {Fasano}, {Popesso}, {Saracco},
  \& {Zaggia}}]{2004A&A...423..761V}
{Vanzella} E. {et~al.}, 2004, \aap, 423, 761

\bibitem[{Wolpert(1992)}]{Wolpert92stackedgeneralization}
Wolpert D.~H., 1992, Neural Networks, 5, 241

\bibitem[{{Yeche} {et~al}\mbox{.}(2009){Yeche}, {Petitjean}, {Rich}, {Aubourg},
  {Busca}, {Hamilton}, {Le Goff}, {Paris}, {Peirani}, {Pichon}, {Rollinde}, \&
  {Vargas-Magana}}]{2009arXiv0910.3770Y}
{Yeche} C. {et~al.}, 2009, ArXiv:0910.3770

\end{thebibliography}

\end{document}